\documentclass[final,onefignum,onetabnum]{siamart171218}

\usepackage{lipsum}
\usepackage{amsfonts}
\usepackage{graphicx}
\usepackage{epstopdf}
\usepackage{algorithmic}
\ifpdf
  \DeclareGraphicsExtensions{.eps,.pdf,.png,.jpg}
\else
  \DeclareGraphicsExtensions{.eps}
\fi

\newsiamremark{remark}{Remark}
\newsiamremark{hypothesis}{Hypothesis}
\crefname{hypothesis}{Hypothesis}{Hypotheses}
\newsiamthm{claim}{Claim}

\headers{ESPINNs for phase-field models}{Callum Marsh, Radek Erban, Andreas Münch}

\title{Extended pseudo-spectral physics-informed neural networks for phase-field models\thanks{Submitted to the {\it SIAM Journal on Scientific Computing} (\today) %Submitted to the editors DATE.
\funding{This paper was funded by The Martingale Foundation. For the purpose of open access, the author has applied a CC BY public copyright licence to any author accepted manuscript arising from this submission.}
}}

\author{Callum Marsh\thanks{Mathematical Institute, University of Oxford, Radcliffe Observatory Quarter, Woodstock Road, Oxford OX2 6GG, United Kingdom
(\email{callum.marsh@maths.ox.ac.uk},
\email{radek.erban@maths.ox.ac.uk},
\email{andreas.muench@maths.ox.ac.uk}).}
  \and Radek Erban\footnotemark[2]
  \and Andreas Münch\footnotemark[2]}

\usepackage{amsopn}

\makeatletter
\newcommand*{\addFileDependency}[1]{% argument=file name and extension
  \typeout{(#1)}% latexmk will find this if $recorder=0 (however, in that case, it will ignore #1 if it is a .aux or .pdf file etc and it exists! if it doesn't exist, it will appear in the list of dependents regardless)
  \@addtofilelist{#1}% if you want it to appear in \listfiles, not really necessary and latexmk doesn't use this
  \IfFileExists{#1}{}{\typeout{No file #1.}}% latexmk will find this message if #1 doesn't exist (yet)
}
\makeatother

\newcommand*{\myexternaldocument}[1]{%
    \externaldocument{#1}%
    \addFileDependency{#1.tex}%
    \addFileDependency{#1.aux}%
}
\ifpdf
\hypersetup{
  pdftitle={Extended SPINNS Method},
  pdfauthor={Callum Marsh, Andreas Münch, Radek Erban}
}
\fi

\myexternaldocument{nnfe_supplement}

\usepackage{subcaption}
\usepackage{wrapfig}
\usepackage{amsmath}
\usepackage{diagbox}

\usepackage{amssymb}
\usepackage{mathtools}
\usepackage[utf8]{inputenc}
\usepackage[makeroom]{cancel}
\usepackage{color}
\usepackage{tikz}
\usepackage{pgfplots}
\pgfplotsset{compat=1.18}
\usepackage{setspace}
\usepackage{graphicx}
\usepackage{diagbox}
\usepackage{bbding}
\usepackage{algorithm}
\usepackage{algorithmic}
\usepackage{hyperref}
\usetikzlibrary{positioning,arrows.meta}

\definecolor{BrickRed}{rgb}{0.8, 0.25, 0.33}
\definecolor{grey}{rgb}{0.5,0.5,0.5}
\definecolor{lightgrey}{rgb}{0.83,0.83,0.83}
\definecolor{darkgrey}{rgb}{0.25,0.25,0.25}

\begin{document}
\maketitle

% REQUIRED
\begin{abstract}
    Phase-field models play a central role in the continuum description of phase separation, in which the bulk free-energy density and the interfacial thickness parameter determine pattern formation and microstructural evolution. In practice, these constitutive quantities are rarely known a priori and must be inferred from limited dynamical observations. In this work, an extended pseudo-spectral physics-informed neural network (ESPINN) framework is developed for the inverse identification of phase-field models from transient snapshot data. It enables the simultaneous recovery of both the bulk chemical potential and unknown gradient coefficients. Numerical experiments on the one-dimensional Cahn–Hilliard equation demonstrate accurate and statistically stable reconstruction in the noiseless regime, with substantial constitutive information recoverable from even a single snapshot pair. In the presence of noise, reconstruction accuracy degrades gracefully, and increasing the number of snapshots improves robustness by reducing variance across runs. These results establish ESPINN as a data-efficient and physically consistent approach for learning free-energy structure in continuum models of phase separation.
\end{abstract}

% REQUIRED
\begin{keywords}
phase separation, phase-field models, Cahn-Hilliard equation, inverse problem, free-energy identification, physics-informed neural networks, pseudo-spectral methods
\end{keywords}

% REQUIRED
\begin{AMS}
35R30, 68T07, 35K55
\end{AMS}

%\clearpage

\section{Introduction}
\label{sec: intro}

Phase separation is a fundamental physical mechanism underlying pattern formation in a wide range of material systems. Classic examples include alloys~\cite{MANZONI2013212, Fratzl} and polymer mixtures~\cite{RODRIGUEZHERNANDEZ20151, BONGIOVANNI2016213}, where phase separation has been studied extensively using continuum models such as the Cahn-Hilliard equation~\cite{Cahn:1958:FEN, cahn_spinodal_1961, hillertSolidsolutionModel1961}. More recently, liquid-liquid phase separation has emerged as a key organising principle in cell biology, where it has been proposed as a mechanism for the formation of membraneless organelles composed of proteins and nucleic acids~\cite{brangwynneGermlineGranulesAre2009}. In these biological systems, the spatial organisation, internal microstructure, and dynamical behaviour of the resulting phases are closely linked to cellular function. 

A central theoretical framework for describing phase separation across such systems is provided by phase-field models. These models introduce using one or more continuous order parameters that evolve along a direction of descent of the free energy functional. They can describe both spatially heterogeneous equilibrium states and the transient dynamics of phase separation. The Cahn-Hilliard model was introduced for non-uniform binary mixtures and has since been extended to a wide variety of settings~\cite{Cahn:1958:FEN, cahn_spinodal_1961, hillertSolidsolutionModel1961}. Related models include the Flory-Huggins type free energies for polymer solutions~\cite{floryThermodynamicsHigh1942, hugginsSolutionsLongChain1941}, as well as extensions that include additional physical effects such as steric constraints or long-range interactions, for example, in Ohta-Kawasaki models for diblock copolymers~\cite{ohta_equilibrium_1986,kawasaki_equilibrium_1988}. In biological contexts, Flory-Huggins polymer solution models and their generalisations, including charged mixture models such as the Voorn-Overbeek model~\cite{overbeekPhaseSeparationPolyelectrolyte1957}, have provided an important theoretical basis for studying biomolecular phase separation.

Despite their widespread use, the predictive capability of phase field models depends critically on the specification of constitutive inputs, in particular the homogeneous (or bulk) free energy density, the interfacial thickness parameter and the mobility coefficient. Among these, the homogeneous free energy density and the interfacial thickness parameter play a central role in determining phase behaviour and microstructure, and will therefore form the primary focus of this paper. These quantities encode information about microscopic interactions and ultimately determine the locations of phase boundaries, interfacial structures, and characteristic length scales of the emerging macrostructure. In materials science, free energy functions are often inferred from extensive experimental measurements using approaches such as CALPHAD~\cite{spencerBriefHistory2008}. However, the effort required to acquire experimental data across the parameter space of interest is substantial or may be unavailable, for example, when designing materials. For biological macromolecules, free energy is highly sensitive to microscopic structure and varies with protein amino acid sequence. 

The problem of inferring free energy structure from simulation data or experimental observations of phase evolution has motivated a broad range of inverse approaches. Classical methods based on optimisation and inverse-problem theory have been applied to phase-field models with varying degrees of success. Zhao et al.~\cite{zhaoLearningPhysicsPattern2020, zhaoImageInversion2021} identify the bulk chemical potential (the derivative of the homogeneous free energy density) and non-linear mobility in Cahn–Hilliard and Allen–Cahn models from a small number of noisy snapshots of pattern evolution. In that setting, treating the mobility, bulk chemical potential and interfacial thickness parameter as unknowns leads to inherent non-identifiability of the inverse problem. Due to scaling degeneracy, these are not uniquely recovered independently. Only particular combinations that determine the dynamics are identifiable. In the present work, we fix the mobility and focus on identifying the remaining constitutive quantities, which can be inferred from transient dynamics but are not recoverable from equilibrium profiles alone, since a rescaling of space can eliminate the interfacial thickness parameter. Related optimisation-based inference approaches are studied by Glasner~\cite{glasnerOptimizationAlgorithms2021}, who considers parameter identification for parabolic partial differential equations (PDEs), including phase-field models for tri-phase diblock copolymer systems.

Against this background, recent work has increasingly explored whether data-driven methods can be used to infer constitutive structure in phase-field models. Recently, attention has been directed towards using neural networks to learn the phase separation in material systems. This development has been driven in particular by the introduction of physics-informed neural networks (PINNs) by Raissi et al.~\cite{raissiPhysicsinformedNeuralNetworks2019}. In this framework, the governing physical laws, typically expressed as PDEs, are incorporated directly into the training objective~\cite{Erban:2026:NNL}. Raissi et al.~\cite{raissiPhysicsinformedNeuralNetworks2019} demonstrate the viability of this approach for both forward prediction of system dynamics and inverse identification of unknown model components from data. An example of the forward problem is provided by An et al.~\cite{anPredictionMicrostructural2025}, who encode the density and chemical potential fields of a Cahn–Hilliard type model for a polymer blend within a neural network. By enforcing the governing equations and initial and boundary conditions through the loss function, the trained network can predict the subsequent evolution of the system. Compared to purely data-driven neural network approaches that aim to learn solutions directly from observations, PINNs retain explicit physical constraints, which can reduce data requirements and mitigate overfitting when training data are limited. An extension of the PINN framework to a spectral discretisation has led to the development of pseudo-spectral PINNs (SPINNs). Using this approach, Zhao~\cite{zhao2021discoveringphasefieldmodels} identified the bulk chemical potential for the Cahn–Hilliard and Allen–Cahn equations, similar to earlier inverse studies~\cite{zhaoLearningPhysicsPattern2020, zhaoImageInversion2021}, from noiseless simulation data, while fixing both the mobility and the interfacial thickness parameter a priori.

In this paper, we present an extension to the SPINN method proposed by Zhao~\cite{zhao2021discoveringphasefieldmodels} for the inverse identification of constitutive quantities of a phase-field model from limited observations of the evolving structure. We now assume that the coefficients of the gradient terms (linear or non-linear) are unknown. For the Cahn-Hilliard and Allen-Cahn equations, this amounts to an unknown interfacial thickness parameter $\varepsilon$. In this paper, we extend the SPINN framework to learn all of these parameters, as well as the bulk chemical potential from dynamically evolving snapshot data. To this end, we perform a systematic numerical investigation of architectural choices, optimisation strategies, and, most importantly, the framework's robustness to noise.

The paper is organised as follows. In Section~\ref{sec: phase-field models}, we introduce phase-field models and highlight two key models: the Cahn-Hilliard equation and Allen-Cahn equation. In Section~\ref{sec: extended spinn framework}, we briefly cover multilayer perceptrons (MLPs), a scalar trainable parameter, and previous work in PINNs before introducing the extended pseudo-spectral physics-informed neural networks (ESPINN) framework for retrieving the free energy function and physical parameters from seen data. In Section~\ref{sec: numerical experiments}, we systematically assess how the ESPINN framework performs on smooth and noisy data, and investigate its sensitivity to architectural and optimisation hyper-parameters, including the learning rate. We finish in Section~\ref{sec: conclusions} by concluding that the ESPINN framework performs remarkably well on data that is not too noisy and is very flexible in architectural choices and optimisation strategies. We also discuss the limitations of the numerical tests and provide suggestions for further investigation.

\section{Phase-field models}
\label{sec: phase-field models}

We consider phase-field models of the form
\begin{equation}
\frac{\partial \phi}{\partial t} \,=\, \mathcal{G} \,\frac{\delta E}{\delta \phi}\,, \qquad \mbox{where} \; \phi = \phi(\boldsymbol{x},t), \; \mbox{for} \;\boldsymbol{x} \in \Omega \; \mbox{and} \; t \in [0,T]\,.
\label{phasefieldmod}
\end{equation}
$\delta E / \delta \phi$ is the functional derivative of the Helmholtz free energy, $E$, with respect to the order parameter, $\phi$ and $\Omega \subset \mathbb{R}^d$ with spatial dimension $d$ while $\mathcal{G}$ is a differential operator. The Helmholtz free energy has the form
\begin{equation*}
E = \int_{\Omega} \left[ \frac{\varepsilon^2}{2} \left| \nabla \phi \right|^2  + F(\phi) \right] \mbox{d}\boldsymbol{x}\,,
\end{equation*}
where the first term is the gradient energy (penalising sharp spatial variations using an interfacial thickness parameter $\varepsilon$) and $F(\phi)$ is the homogeneous free energy density. Therefore, we have that
\begin{equation*}
\frac{\delta E}{\delta \phi} = f(\phi) - \varepsilon^2 \Delta \phi,
\end{equation*}
with $f(\phi) = F'(\phi)$ being the bulk chemical potential. In our illustrative simulations, the homogeneous free energy density $F(\phi)$ will be chosen to be either a symmetric double well (polynomial free energy)
\begin{equation}
F(\phi) = \frac{(\phi^2 - 1)^2}{4}\,, 
\label{eqhf1}
\end{equation}
or logarithmic free energy \cite{Cahn:1958:FEN}
\begin{equation}
F(\phi) = \phi\log(\phi) \,+\, (1-\phi) \log(1-\phi) \,+\, \frac{5 \,\phi\,(1-\phi)}{2}\,, \label{eqhf2}
\end{equation}
with both choices~(\ref{eqhf1}) and (\ref{eqhf2}) having two stable equilibrium phases. The homogeneous free energy (\ref{eqhf1}) has two minima at $\pm1$ and a maximum at $\phi=0$ while the logarithmic form (\ref{eqhf2}) has a maximum at $\phi = 0.5$ with minima at $\phi \approx 0.145$ and $\phi \approx 0.855$. 

Possibly the simplest example of~(\ref{phasefieldmod}) is the Allen-Cahn equation in which we let $ \mathcal{G} = -M$ with constant mobility $M > 0$, resulting in
\begin{equation}
\label{allencahn}
\frac{\partial \phi}{\partial t} \,=\, M \big( \varepsilon^2 \Delta \phi - f(\phi) \big)\,.
\end{equation} 
This is a fundamental phase-field model used to describe the evolution of interfaces in systems that undergo phase transitions, such as the transformation between ice and liquid water. In this framework, the state of the material is represented by an order parameter, denoted here by $\phi(\boldsymbol{x}, t)$, which serves as a continuous indicator of the local phase. In the case of ice-water transitions, if using the polynomial free energy~(\ref{eqhf1}) as a modelling assumption, we could assign $\phi = 1$ to denote regions of solid ice and $\phi = -1$ to regions of liquid water. The intermediate values $-1 < \phi < 1$ correspond to the diffuse interface, where the material undergoes a continuous transition between the two phases. Since our research focuses on phase separation, we will primarily study the Cahn-Hilliard equation. In this case, we have $ \mathcal{G} = M \Delta $ in equation~(\ref{phasefieldmod}), with $M > 0$ a constant, and so
\begin{equation}
\label{cahnhilliardeq}
\frac{\partial \phi}{\partial t} \,=\, M \Delta \big( f(\phi) - \varepsilon^2 \Delta \phi \big)\,.
\end{equation}
Using the Cahn-Hilliard equation, we provide a thermodynamically consistent framework for describing phase separation in systems where the order parameter is conserved, such as the segregation of biomolecular components within a biological cell. In~this context, the order parameter $\phi(\boldsymbol{x}, t)$ represents a normalised local concentration difference of two chemical species or phases, say chemical I and chemical II. Essentially, in the case of assuming a polynomial free energy~(\ref{eqhf1}), the regions where $\phi = 1$ represent regions populated solely by chemical I, while regions with $\phi = -1$ represent regions populated solely by chemical II. From an initially perturbed homogenous mixture, the temporal evolution of $\phi$ follows the Cahn-Hilliard equation~(\ref{cahnhilliardeq}), which captures how small fluctuations in the concentration can spontaneously amplify when the mixture is thermodynamically unstable. This leads to phase separation into dense regions of chemical I and chemical II, respectively.

Both the Allen-Cahn equation~(\ref{allencahn}) and the Cahn-Hilliard equation~(\ref{cahnhilliardeq}) can be cast within a unified gradient-flow framework. Recall that the evolution of the order parameter $\phi(\boldsymbol{x},t)$ is given by \eqref{phasefieldmod}. Assuming that $\mathcal{G}$ is self-adjoint and negative semidefinite and imposing the vanishing boundary contributions, the time derivative of the Helmholtz free energy satisfies
\begin{equation*}
\frac{\mbox{d}E}{\mbox{d}t} 
= \int_\Omega \frac{\delta E}{\delta \phi} \, \mathcal{G} \frac{\delta E}{\delta \phi} \, \mbox{d} \boldsymbol{x} \leq 0\,.
\end{equation*}
Stationary states occur when we have equality, which corresponds to $\delta E/\delta \phi \equiv 0$ for the Allen-Cahn equation~(\ref{allencahn}) and $\nabla \delta E/\delta \phi \equiv 0$ for the Cahn-Hilliard equation~(\ref{cahnhilliardeq}).

\begin{figure}[tb]
\centering
\begin{subfigure}{0.49\textwidth}
        \centering
        \includegraphics[width=\linewidth]{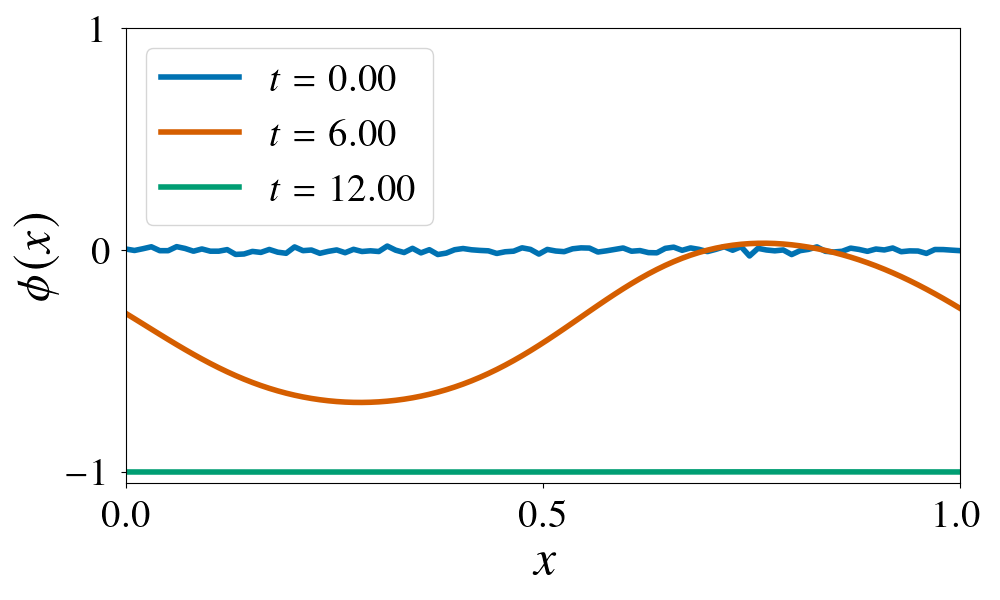}
        \caption{Allen-Cahn}
        \label{subfig: evolution graphs - AC}
\end{subfigure}
\hfill
\begin{subfigure}{0.49\textwidth}
        \centering
        \includegraphics[width=\linewidth]{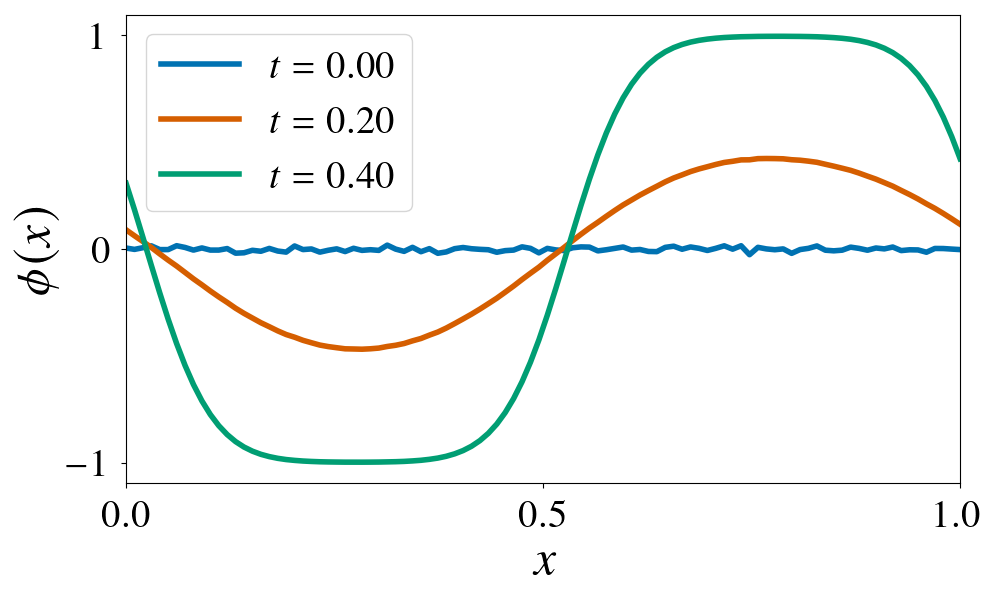}
        \caption{Cahn-Hilliard}
        \label{subfig: evolution graphs - CH}
\end{subfigure}
\caption[Time evolution of phase-field equations]{{\it Time evolution of the Allen-Cahn equation~$(\ref{allencahn})$ and Cahn-Hilliard equation~$(\ref{cahnhilliardeq})$ from identical initial conditions in one spatial dimension, i.e. we have $d=1$ with $\Omega=[0,1]$. Both simulations use time step $\mbox{{\rm d}}t = 10^{-4}$, space discretization $\mbox{{\rm d}}x = 10^{-2}$, $\varepsilon = 0.05$ and $f(\phi) = \phi^3 - \phi$.}}
\label{fig: evolution graphs}
\end{figure}

Figure~\ref{fig: evolution graphs} shows snapshots from the time evolution of one-dimensional simulations (i.e., $d=1$ and $\Omega=[0,1]$) of the Allen-Cahn equation~(\ref{allencahn}) and the Cahn-Hilliard~(\ref{cahnhilliardeq}) equation, starting from identical initial conditions. Although the two models share the same free-energy functional, their dynamics differ markedly. First, the Allen-Cahn solution evolves more slowly than the Cahn-Hilliard solution. Since Allen–Cahn dynamics do not conserve mass, interfaces are not constrained to persist, and the solution gradually relaxes toward a single uniform phase without the formation of sharp transition layers, as shown in Figure \ref{subfig: evolution graphs - AC}. In contrast, the Cahn–Hilliard solution undergoes rapid phase separation, as evidenced by the emergence of steep interfaces and extended plateaus near the stable states in Figure \ref{subfig: evolution graphs - CH}. Mass conservation enforces phase coexistence, while gradients in the chemical potential drive sustained interfacial motion and coarsening. Consequently, the Cahn–Hilliard dynamics exhibit a more pronounced spatial structure and faster evolution than the Allen–Cahn dynamics over the same time interval.

\section{Extended SPINN framework}
\label{sec: extended spinn framework} In this section, we first introduce the background theory of feed-forward neural networks, i.e. networks that map inputs to outputs by propagating information through a sequence of weighted layers without feedback or recurrence~\cite{Goodfellow:2016:DL}. We then present the ESPINN framework in Section~\ref{subsec: extended SPINN method}.

\subsection{Multilayer perceptrons (MLPs)}
\label{subsecMLPs} MLPs are feed-forward neural networks where adjacent layers are fully connected, each with non-linear activation functions. Notably, these neurons are organised into layers, comprising an input layer, hidden layers, and an output layer. The depth of an MLP is defined as the number of hidden layers. Given an input ${\boldsymbol{x}} \in \mathbb{R}^{n_1}$, then, for each subsequent layer, we define $\boldsymbol{a}^{[\ell]} \in \mathbb{R}^{n_\ell}$ as the output of the $\ell$-th layer where $n_\ell$ is the number of neurons in such layer~\cite{Deep}. More formally, an $L$-layer MLP is defined by
\begin{align}
\boldsymbol{a}^{[1]} &= \boldsymbol{x} \in \mathbb{R}^{n_1},
\label{eq: MPL input}\\
\boldsymbol{a}^{[\ell]} &= \boldsymbol{\omega}^{[\ell]} \!\left( \boldsymbol{W}^{[\ell]} \boldsymbol{a}^{[\ell-1]} + \boldsymbol{b}^{[\ell]} \right) \in \mathbb{R}^{n_\ell} , \,\,\,\, \ell = 2, 3, \cdots L,
\label{eq: MLP recursive}
\end{align}
where $\boldsymbol{W}^{[\ell]} \in {\mathbb R}^{n_{\ell-1} \times n_\ell}$ and $\boldsymbol{b}^{[\ell]}$ are the weights and biases of the $\ell$-th layer, respectively, and $\boldsymbol{\omega}^{[\ell]} : {\mathbb R}^{n_\ell} \to {\mathbb R}^{n_\ell}$ is given by
\begin{equation}
\boldsymbol{\omega}^{[\ell]}{(\mathbf z})
=
[\omega(z_1), \omega(z_2), \dots, \omega(z_{n_\ell})]\,,
\qquad
\mbox{for} \; {\mathbf z} = [z_1,z_2,\dots,z_{n_\ell}]\,.
\label{vecactfun}
\end{equation}
$\omega: {\mathbb R} \to {\mathbb R}$ is the activation function. In our computational explorations, we test the following activation functions $\omega$:

\begin{align*}
\text{sigmoid}(z) &= \sigma(z) = \frac{e^{z}}{1 + e^{z}}, & \operatorname{Tanh}(z) &= \frac{e^{z} - e^{-z}}{e^{z} + e^{-z}}, \\
\operatorname{ReLU}(z) &= \begin{cases}
z, & z > 0,\\
0, & z \le 0,
\end{cases}
&
\operatorname{SiLU}(z) &= z \cdot \sigma(z). \\
\operatorname{ELU}(z) &= \begin{cases}
z, & z > 0,\\
e^{z} - 1, & z \le 0,
\end{cases}
&
\operatorname{GELU}(z) &= \frac{z}{2}\!\left[1 + \operatorname{erf}\!\left(\frac{z}{\sqrt{2}}\right)\right].
\end{align*}

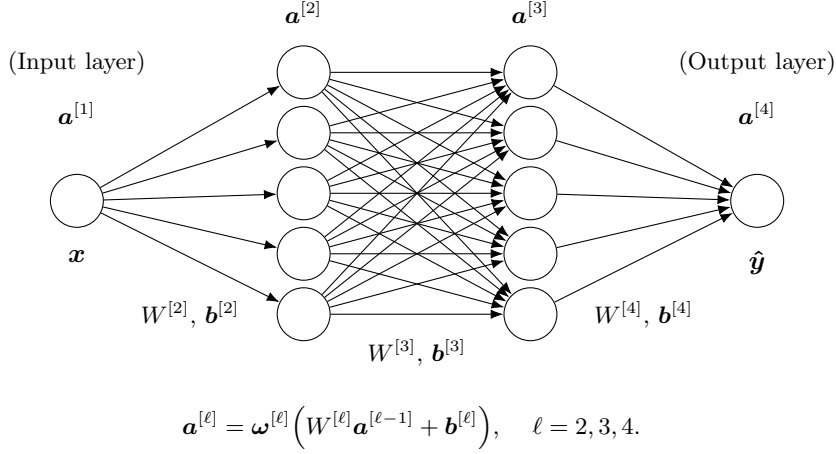
\begin{figure}[tb]
    \centering
    \label{fig: MLP schematic}
    \begin{tikzpicture}[
      neuron/.style={circle,draw,minimum size=7mm,inner sep=0pt},
      layerlabel/.style={font=\small},
      >=Latex
    ]
    
    % Input layer (a^[1])
    \node[layerlabel, align=center] at (0,1.5) {(Input layer) \\[10pt] $\boldsymbol{a}^{[1]}$};
    \node[neuron] (I) at (0,0) {};
    \node[below=5pt of I] {$\boldsymbol{x}$};
    
    % Hidden layer 1 (a^[2])
    \node[layerlabel] at (3,2.5) {$\boldsymbol{a}^{[2]}$};
    \foreach \i in {1,...,5}{
      \node[neuron] (H1-\i) at (3,2.5-0.8*\i) {};
    }
    
    % Hidden layer 2 (a^[3])
    \node[layerlabel] at (6,2.5) {$\boldsymbol{a}^{[3]}$};
    \foreach \i in {1,...,5}{
      \node[neuron] (H2-\i) at (6,2.5-0.8*\i) {};
    }
    
    % Output layer (a^[4])
    \node[layerlabel, align=center] at (9,1.5) {(Output layer) \\[10pt] $\boldsymbol{a}^{[4]}$};
    \node[neuron] (O) at (9,0) {};
    \node[below=5pt of O] {$\boldsymbol{\hat{y}}$};
    
    % Connections
    \foreach \j in {1,...,5}{
      \draw[->,thin] (I) -- (H1-\j);
    }
    \foreach \i in {1,...,5}{
      \foreach \j in {1,...,5}{
        \draw[->,thin] (H1-\i) -- (H2-\j);
      }
    }
    \foreach \i in {1,...,5}{
      \draw[->,thin] (H2-\i) -- (O);
    }
    
    % Weight and bias annotations
    \node[font=\small] at (1.5,-1.5) {$W^{[2]},\,\boldsymbol{b}^{[2]}$};
    \node[font=\small] at (4.5,-2.0) {$W^{[3]},\,\boldsymbol{b}^{[3]}$};
    \node[font=\small] at (7.5,-1.5) {$W^{[4]},\,\boldsymbol{b}^{[4]}$};
    
    % Equation annotation
    \node[font=\small,align=left] at (4.5,-3.0) {
      $\boldsymbol{a}^{[\ell]} = \boldsymbol{\omega}^{[\ell]}\!\left(W^{[\ell]} \boldsymbol{a}^{[\ell-1]} + \boldsymbol{b}^{[\ell]}\right)$, \quad $\ell = 2,3,4$.
    };
    
\end{tikzpicture}
\caption[Multilayer Perceptron (MLP) schematic]{{\it Schematic example of an MLP with a single input-output and two hidden layers with five neurons in each layer.}}
\end{figure}

\subsection{Trainable scalar parameter}
\label{subsec: single neuron approximator}
\begin{figure}[tb]
\centering
    \begin{tikzpicture}
    \begin{axis}[
        width=0.8\linewidth,
        height=5cm,
        axis lines=middle,
        xlabel={$\theta$},
        ylabel={$\gamma$},
        xmin=-6, xmax=6,
        ymin=0, ymax=1,
        samples=200,
        ticks=none
    ]
    % Rescaled sigmoid
    \addplot[thick, blue] {0.2 + 0.6*(1/(1+exp(-x)))};
    
    % Bounds
    \addplot[dashed] coordinates {(-5,0.2) (6,0.2)};
    \addplot[dashed] coordinates {(-5,0.8) (6,0.8)};
    
    % Theta_0
    \addplot[dashed] coordinates {(0,0.2) (0,0.5)};
    \addplot[mark=*, only marks] coordinates {(0,0.5)};
    
    \node[left] at (axis cs:-5,0.2) {$\gamma_{\mathrm{lb}}$};
    \node[left] at (axis cs:-5,0.8) {$\gamma_{\mathrm{ub}}$};
    \node[above right] at (axis cs:-0.7,0.5) {$\gamma_0$};
    
    \end{axis}
\end{tikzpicture}
\caption[Transformed sigmoid function]{{\it Sigmoid reparameterisation used for bounded scalar parameters. Initialisation at $\theta_{0} = 0$ places the parameter in the linear regime of the sigmoid, ensuring non-vanishing gradients during early optimisation.}}
\label{fig:sigmoid_param}
\end{figure}
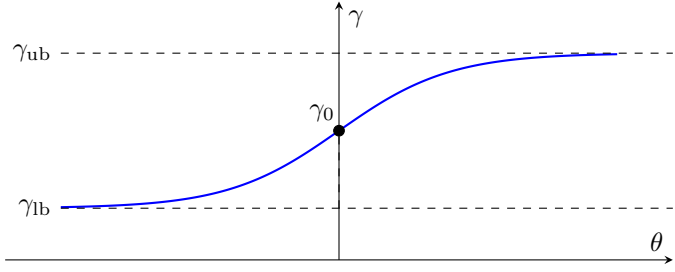

As shown in Figure \ref{fig:sigmoid_param}, to represent an unknown coefficient $\gamma \in \mathbb{R}$, we parametrise it by a single trainable scalar $\theta$ and a sigmoid based reparametrisation that enforces the bounds $\gamma_{lb}$ and $\gamma_{ub}$:
\begin{equation}
\label{eq: parameter estimator}
    \gamma = \gamma_{lb} + (\gamma_{ub} - \gamma_{lb}) \sigma(\theta).
\end{equation}
The coefficient $\theta_\gamma$ is optimised concurrently with the network for $f(\phi)$. This joint optimisation allows the parameter to be learned through the same gradient-based updates that drive the training of $f(\phi)$, ensuring that both components adapt consistently to the data. The derivative that the optimiser sees is
\begin{equation*}
    \frac{d\gamma}{d\theta} = (\gamma_{ub} - \gamma_{lb}) \sigma(\theta)(1 - \sigma(\theta))
\end{equation*}
so the gradient magnitudes are proportional to the interval width $(\gamma_{ub} - \gamma_{lb})$ and are largest when $\sigma(\theta) \approx 0.5$. In practice, we start with a physically reasonable prior guess $\gamma_0$. From the mapping
\begin{equation*}
    \gamma_0 = \gamma_{lb} + (\gamma_{ub} - \gamma_{lb}) \sigma(\theta_0),
\end{equation*}
we solve for the initial $\theta_0$:
\begin{equation*}
    \theta_0 = \ln \frac{p}{1-p} \,\,\,\,\,\,\,\,\,\, \text{where} \,\,\,\,\,\,\,\,\,\, p = \frac{\gamma_0 - \gamma_{lb}}{\gamma_{ub} - \gamma_{lb}}.
\end{equation*}
For our simulations, we choose $\gamma_0$ to lie in the middle of the range $[\gamma_{lb}, \gamma_{ub}]$. We therefore have $p=0.5$ and so $\theta_0 = 0$, precisely the linear region of the sigmoid function giving strong, non-vanishing gradients.

\subsection{Physics-informed Neural Networks}

The Physics-Informed Neural Network (PINN) framework, initially proposed by Raissi et al. \cite{raissiPhysicsinformedNeuralNetworks2019}, provides a flexible, mesh-free approach to solving PDEs by embedding the underlying physical laws directly into the loss function of a neural network. In this approach, a deep neural network $\boldsymbol{u}_\Theta(\boldsymbol{x}, t)$ is parametrised by weights $\Theta$ and serves as a surrogate model for the unknown solution of the governing equation.

The training process of a PINN involves minimising a composite loss function that balances the network's accuracy on known data with its adherence to the physical model. For a PDE of the general form
\begin{equation*}
%\label{eq: PDE pinns}
\boldsymbol{u}_t + \mathcal{N}[\boldsymbol{u}] = 0, \quad t \in [0, T] \quad \boldsymbol{x} \in \Omega,
\end{equation*}
with initial and boundary conditions
\begin{align*}
    \boldsymbol{u}(\boldsymbol{x}, t) &= \boldsymbol{g}(\boldsymbol{x}) \quad \boldsymbol{x} \in \Omega \\
    \mathcal{B}[\boldsymbol{u}] &= 0, \quad t \in [0, T] \quad \boldsymbol{x} \in \partial \Omega,
\end{align*}
the unknown solution $\boldsymbol{u}(\boldsymbol{x}, t)$ is represented by the neural network $\boldsymbol{u}_{\Theta}(\boldsymbol{x}, t)$, where $\Theta$ represents the hyper parameters of the network. Using weighted loss function formulation \cite{cai2021physicsinformedneuralnetworkspinns, cuomo2022scientific, wang2022respectingcausalityneedtraining}, the goal is to train the subsequent model by minimising
\begin{equation*}
    \mathcal{L}(\Theta)
    =
    \lambda_{ic}\,\mathcal{L}_{ic}(\Theta)
    +
    \lambda_{bc}\,\mathcal{L}_{bc}(\Theta)
    +
    \lambda_{r}\,\mathcal{L}_{r}(\Theta),
%    \label{eq:pinn_loss}
\end{equation*}
where
\begin{equation*}
    \mathcal{L}_{ic}(\Theta)
    =
    \frac{1}{N_{ic}}
    \sum_{i=1}^{N_{ic}}
    \left|
        \boldsymbol{u}_{\Theta}(\boldsymbol{x}^{i}_{ic}, 0) - \boldsymbol{g}(\boldsymbol{x}^{i}_{ic})
    \right|^{2},
%    \label{eq:pinn_ic}
\end{equation*}
\begin{equation*}
    \mathcal{L}_{bc}(\Theta)
    =
    \frac{1}{N_{bc}}
    \sum_{i=1}^{N_{bc}}
    \left|
        \mathcal{B}\!\left[\boldsymbol{u}_{\Theta}\right]\!(\boldsymbol{x}^{i}_{bc}, t^{i}_{bc})
    \right|^{2},
%    \label{eq:pinn_bc}
\end{equation*}
and
\begin{equation*}
    \mathcal{L}_{r}(\Theta)
    =
    \frac{1}{N_{r}}
    \sum_{i=1}^{N_{r}}
    \left|
        \frac{\partial \boldsymbol{u}_{\Theta}}{\partial t}(\boldsymbol{x}^{i}_{r}, t^{i}_{r})
        +
        \mathcal{N}[\boldsymbol{u}_{\Theta}](\boldsymbol{x}^{i}_{r}, t^{i}_{r})
    \right|^{2}.
%    \label{eq:pinn_residual}
\end{equation*}

The latter is computed by applying automatic differentiation to the neural network output, thereby enabling efficient evaluation of derivatives of arbitrary order. PINNs are trained using standard stochastic gradient descent or adaptive optimisation algorithms such as Adam. Rathore et al. \cite{rathore2024challengestrainingpinnsloss} showed that the combination of Adam training followed by the quasi-Newton method L-BFGS was found to be superior compared to using Adam or L-BFGS alone.  Collocation points are sampled in the domain to evaluate the PDE residual, while separate points on the boundary and initial surfaces are used to enforce the boundary and initial conditions. This approach allows PINNs to generalise across complex geometries and high-dimensional problems without requiring an explicit mesh or discretisation scheme.

Zhao~\cite{zhao2021discoveringphasefieldmodels} introduced the SPINN framework, designed to directly infer the bulk chemical potential $f(\phi)$ of phase-field models from image data, without the need for densely sampled spatio-temporal datasets. The method combines the approximation capabilities of PINNs with the computational efficiency of pseudo-spectral discretisation schemes. A significant feature is that instead of using a neural network to approximate the field variables, the non-linear bulk chemical potential $f(\phi)$ is approximated by a neural network. A limitation of Zhao's method is that model parameters, such as the interfacial thickness parameter, must be known to apply it. We now propose a Pseudo-spectral framework in which these physical parameters are unknown and can be simultaneously learned with the free energy function.

\subsection{Extended SPINN methodology}
\label{subsec: extended SPINN method}

We now propose a problem in which we have the data from a phase-field simulation. We assume we know the general form of the data-generating equation with constant mobility. Beyond that, the free energy function and other physical parameters are unknown. Now, our goal is to discover the bulk chemical potential $f(\phi)$ and parameters associated with the derivatives of $\phi$ within a phase-field system. This problem arose from the goal of deriving constitutive properties of coarse-grained systems from stochastic systems that exhibit behaviour similar to that of a phase-field model. In these systems, we do not know the associated parameters and are essentially proposing that they describe the same underlying physics as a phase-field model.

Consider a general phase-field PDE problem for $\phi(\boldsymbol{x},t)$ of the form
\begin{equation}
\label{eq: general phase field}
    \frac{\partial \phi}{\partial t} = \mathcal{G} \left[ \sum_j \mu_j \, g_j \left(\phi, \nabla \phi, \Delta \phi \right) + f(\phi) \right], \qquad \mbox{for} \; (\boldsymbol{x},t) \in \Omega \times (0,T]
\end{equation}
with periodic boundary conditions. Here, the operators $\mathcal{G}$ and $g_j$ are known, but the function $f$ and parameters $\mu_j$ are unknown. 

Let $\Phi$ be the approximation of $\phi$ on a discrete spatial equidistant grid with a fixed number of grid points. Then, the collected data is in the form 
\begin{equation}
    {\cal{D}} = \left \{ (\Phi_i^{(1)}, \Phi_i^{(2)}, (\Delta t)_i \right \}_{i=1}^N \subset \Omega \times \Omega \times \mathbb{R}^+,
    \label{data_form}
\end{equation}
where we have $N$ pairs of snapshots $(\Phi_i^{(1)}, \Phi_i^{(2)})$ each separated with a time of $(\Delta t)_i$. Similarly, we let $\mathcal{G}_h$ be a matrix operator that approximates the operator $\mathcal{G}$ on the discrete spatial equidistant grid.

By splitting the $\mu_j \, g_j$ terms, we write the spatially discrete form as 
\begin{equation*}
\frac{\partial \Phi}{\partial t} = \mathcal{G}_h \left[  
\sum_j \alpha_j \, L_j(\Phi) + \sum_j \beta_j \, N_j (\Phi) + f(\Phi) \right],
\end{equation*}
where $L_j(\Phi)$ are linear terms and $N_j(\Phi)$ are the rest. $\alpha_j$ and $\beta_j$ represent the parameter coefficients of the linear operators and non-linear terms, respectively. The goal of this section is to use the data of the form \eqref{data_form} to find the function $f$ and the values of the parameters $\alpha_j$ and $\beta_j$. To do this, we require a loss function for training the neural networks. For one time step, $\Delta t$, we use the semi-implicit linear stabilised scheme
\begin{equation*}
\frac{\Phi_{t+\Delta t} - \Phi_t}{\Delta t} = \mathcal{G}_h \left[ \sum_j \alpha_j \, L_j(\Phi_{t+\Delta t}) + \sum_j \beta_j \, N_j (\Phi_t) + f(\Phi_t) + C(\Phi_{t+\Delta t} - \Phi_t) \right].
\end{equation*}
Therefore, our time-stepping scheme looks as follows
\begin{multline*}
    \Phi_{t+ \Delta t} = \left( \! 1 - \Delta t \,\mathcal{G}_h \! \left[C + \sum_j \alpha_i \, L_j \right] \!\right)^{\!\!\!-1}
    \\ \left\{ \Phi_t + \Delta t \, \mathcal{G}_h \! \left[ \sum_j \beta_j \, N_j(\Phi_t) + f(\Phi_t) - C\Phi_t \right] \!\right\}.
\end{multline*}
We define our function and parameter estimators as
\begin{equation*}
\mathcal{N}_f: \Phi \to \mathcal{N}_f(\Phi;\theta_f), \qquad \mathcal{N}_{\alpha_j}(\theta_{\alpha_j}), \qquad  \mathcal{N}_{\beta_j}(\theta_{\beta_j}),
\end{equation*}
where $\mathcal{N}_f(\Phi;\theta_f)$ is an MLP (\ref{eq: MPL input})--(\ref{vecactfun}) meant to reconstruct $f(\Phi)$ while each $\mathcal{N}_{\alpha_j}(\theta_{\alpha_j})$ and $\mathcal{N}_{\beta_j}(\theta_{\beta_j})$ are single parameter estimators \eqref{eq: parameter estimator} for the corresponding  $\alpha_j$ or $\beta_j$. $\theta_k$ represents the free parameters of the corresponding model neural network. We define the extended linear SPINN loss function by:

Given the data $(\Phi^{(1)},\Delta t)$, we define
\begin{multline}
    \mathcal{N}_R: (\Phi^{(1)},\Delta t) \to \left( 1 - \Delta t \mathcal{G}_h \left[C + \sum_j \mathcal{N}_{\alpha_j}(\theta_{\alpha_j}) \, L_j \right] \right)^{-1} \nonumber\\ \left\{ \Phi^{(1)} + \Delta t \mathcal{G}_h \left[ \sum_j \mathcal{N}_{\beta_j}(\theta_{\beta_j}) \, N_j(\Phi^{(1)}) + \mathcal{N}_f(\Phi^{(1)};\theta_f) - C\Phi_t \right] \right\}.
\end{multline}
Then the loss function is defined by
\begin{equation}
\label{eq: loss function}
    \mathcal{L}(\Theta) = \frac{C_0}{N} \sum_{i=1}^N \| \Phi_i^{(2)} - \mathcal{N}_R(\Phi_i^{(1)}, (\Delta t)_i\,\, ; \Theta ) \|_2^2,
\end{equation}
where $\Theta = \{\theta_f, \theta_{\alpha_1}, \cdots , \theta_{\beta_1}, \cdots\}$ is the set of all trainable parameters in all MLPs (\ref{eq: MPL input})--(\ref{vecactfun}) and trainable scalar parameters \eqref{eq: parameter estimator} in $\mathcal{N}_R$.
By minimising this loss function, we can identify $f(\Phi)$ by the neural network $\mathcal{N}_f(\Phi;\theta_j)$ and parameters, $\alpha_j$ and $\beta_j$, by the estimators $\mathcal{N}_{\alpha_j}(\theta_{\alpha_j})$ and $\mathcal{N}_{\beta_j}(\theta_{\beta_j})$, respectively. All networks/estimators are trained simultaneously. The parameter $C_0$ is a key hyperparameter that prevents our loss function from reducing to a point where vanishing gradients prevent further improvement. The choice of $C_0$ is not unique, does not alter the minimiser of the loss, and serves primarily to improve numerical conditioning. In practice, we adjust $C_0$ such that the final loss after training is approximately $\mathcal{O}(10^{-4})$ by performing a small initial test run. The adjustable nature of $C_0$ allows us to optimise the training procedure based on the noise observed in the data. A higher $C_0$ is therefore used on smooth data, while a lower $C_0$ is used on noisy data to reduce the risk of the training overfitting to the noise, especially in the reconstruction of $f(\phi)$. Our goal is to find a deterministic smooth $f(\phi)$ even in noisy data.

We model the bulk chemical potential $f$ using an MLP (as described in \ref{subsecMLPs}). For the approximation of the interfacial thickness parameter $\varepsilon$, we first define $\gamma = \varepsilon^2$, i.e. the gradient coefficient, which is then parameterised by $\mathcal{N}_\gamma(\theta_\gamma)$, a single parameter estimator ( as described in \ref{subsec: single neuron approximator}). For smooth data, we train the networks using Adam \cite{kingma2017adammethodstochasticoptimization} followed by L-BFGS \cite{liu1989limited}, whereas for noisy data, we perform only Adam.

\section{Cahn- Hilliard numerical experiments}
\label{sec: numerical experiments}
To acquire the data, we simulate a 1D Cahn-Hilliard system using a linear semi-implicit pseudo-spectral time stepping method with $x \in [0,1]$, $\Delta x = 10^{-2}$ and $\Delta t = 10^{-4}$ saving the state of $\Phi(x,t)$ at each time step. We use an initial condition of
\begin{equation*}
    \Phi(x,0) = c + \delta \xi(x)
\end{equation*}
where $\delta = 10^{-4}$ with $c=0$ for the polynomial free energy \eqref{eqhf1} and $c=0.5$ for the logarithmic free energy \eqref{eqhf2}. $\xi(x)$ is sampled from a uniform distribution in the range $[-1,1]$ for each $x \in [0,1]$. We run this until the system has reached a stable equilibrium.

We return to the phase-field models of the form \ref{eq: general phase field} and let $M = 1$, since this adjusts the speed at which phase separation occurs (and so is somewhat arbitrary). We introduce the subscript $h$ to denote the discrete spatial operator arising from the chosen spatial discretisation; in this case, we choose 100 equally spaced points in $x \in [0,1]$. By discretising the equations and parametrising the gradient coefficient $\gamma$ as $\mathcal{N}_\gamma(\theta_\gamma)$, we have that
\begin{equation*}
    \sum_j \alpha_j \, L_j(\Phi) = - \mathcal{N}_\gamma(\theta_\gamma) \cdot (\partial_{xx}\Phi)_h, \qquad \sum_j \beta_j \, N_j(\Phi) = 0.
\end{equation*}
while $\mathcal{G}_h = (\partial_{xx})_h$. We used the stabilising term $C=-2(\partial_{xx})_h$. The estimator and stabilising term, along with the the MLP $\mathcal{N}_f(\Phi, \theta_f)$, are then fed into the extended linear SPINN loss function \eqref{eq: loss function} solved with a pseudo-spectral method and trained on $N$ random snapshot pairs from data in the form of \ref{data_form} taken suitably after the initial random state has settled when the problem is in its dynamic stage.

We tested the method on the bulk chemical potentials
\begin{equation*}
f(\phi) = \phi^3 - \phi \qquad\mbox{and}\qquad
f(\phi) = \log(\phi) - \log(1-\phi) + \frac{5}{2} - 5 \phi, 
\end{equation*}
which are derivatives of the homogeneous free energies (\ref{eqhf1}) and (\ref{eqhf2}), respectively. For training, we used Adam for 20,000 epochs, followed by L-BFGS, given that the data are smooth, with a maximum of 3,000 epochs. For our spatial and temporal parameters, we use $dx = 0.01$ and $dt = 10^{-4}$. Due to the size of $dx$, we chose reasonable limits on the gradient coefficient $\gamma$ of 
\begin{equation*}
[\gamma_{lb}, \gamma_{ub}] = [10^{-4}, 0.1].
\end{equation*} 
We report the mean and variance over 100 runs, each initialised with a different seed. For each run, the seed randomises the $N$ snapshots used as the training data, the parameter initialisation $\Theta$, and, in the case of noisy data, the noise applied to the data before training. Such randomisation was chosen to demonstrate the model's robustness.

The model is tested on $\phi$ values in the possible domain of the bulk chemical potential $f(\phi)$. A test run was first performed with $C_0=10^{10}$, which was then adjusted as necessary such that the final loss $\mathcal{L} \sim \mathcal{O}(10^{-4})$. The training is then run using the adjusted $C_0$. To find our overall errors, we first define the outputs of our trained neural network and parameter estimator as
\begin{equation*}
\hat{f} := \mathcal{N}_f(\Phi, \theta_f), \qquad \hat{\varepsilon} := \sqrt{\mathcal{N}_{\gamma}(\theta_{\gamma})}, \qquad \Phi \in [a,b],
\end{equation*}
where $[a,b]$ is our domain space. Furthermore, let the relative errors between the predicted $\hat{f}$ and $\hat{\varepsilon}$ and that of the actual bulk chemical potential $f$ and interfacial thickness parameter $\varepsilon$ as
\begin{equation*}
E_f := \frac{||\hat{f} - f||_{L^2}}{||f||_{L^2}}, \qquad E_{\varepsilon} := \frac{|\hat{\varepsilon} - \varepsilon|}{\varepsilon}
\end{equation*}

In Sections~\ref{subsec: noiseless data}, \ref{subsec: noise robustness} and \ref{subsec: learning rate sensitivity}, we trained the model with the neural network $\mathcal{N}_f$ having 2 layers of 20 neurons each with a SiLU activation function. In Section~\ref{subsec: architectural sensitivity}, we perform extensive testing of the activation functions described in Section~\ref{subsecMLPs} with varying widths and layers to the neural network $\mathcal{N}_f$.

\subsection{Application on noiseless data}
\label{subsec: noiseless data}

Due to the lack of stochasticity, we found that using a large loss function multiplier of $C_0 = 10^{16}$ to be optimal. Once Adam has brought the parameters into a suitable basin, the subsequent application of L-BFGS significantly accelerates convergence and refines the solution. In the absence of noise, the L-BFGS stage is particularly effective, as it exploits the smoothness of the objective landscape to achieve rapid and precise convergence to a low-residual solution. Remarkably, even with only a single snapshot pair $(N=1)$, the predicted chemical potential already closely matches the true function with an almost perfect prediction for the interfacial width parameter. The qualitative structure of $f(\phi)$, including the location of extrema and overall curvature, is recovered. The small shaded regions on either side of the ensemble mean indicate that some variability remains across randomly selected points and initialisations. Increasing to only two snapshot pairs $(N=2)$ leads to a tighter concentration of inferred $\hat{f}(\phi)$ around $f(\phi)$ such that there is no longer any visible variance in the reconstruction. The associated estimates for the interfacial thickness parameter $\varepsilon$ remain closely centred around the true value.

These results suggest that substantial constitutive information is already encoded in even a single evolutionary step and that the Extended SPINN framework can use this to accurately and stably recover the functional and scalar constitutive components. The additional snapshot pair primarily improves statistical stability rather than correcting any systemic bias.

\begin{figure}[tb]
    \centering
    \begin{subfigure}[b]{0.49\textwidth}
        \includegraphics[width=\textwidth]{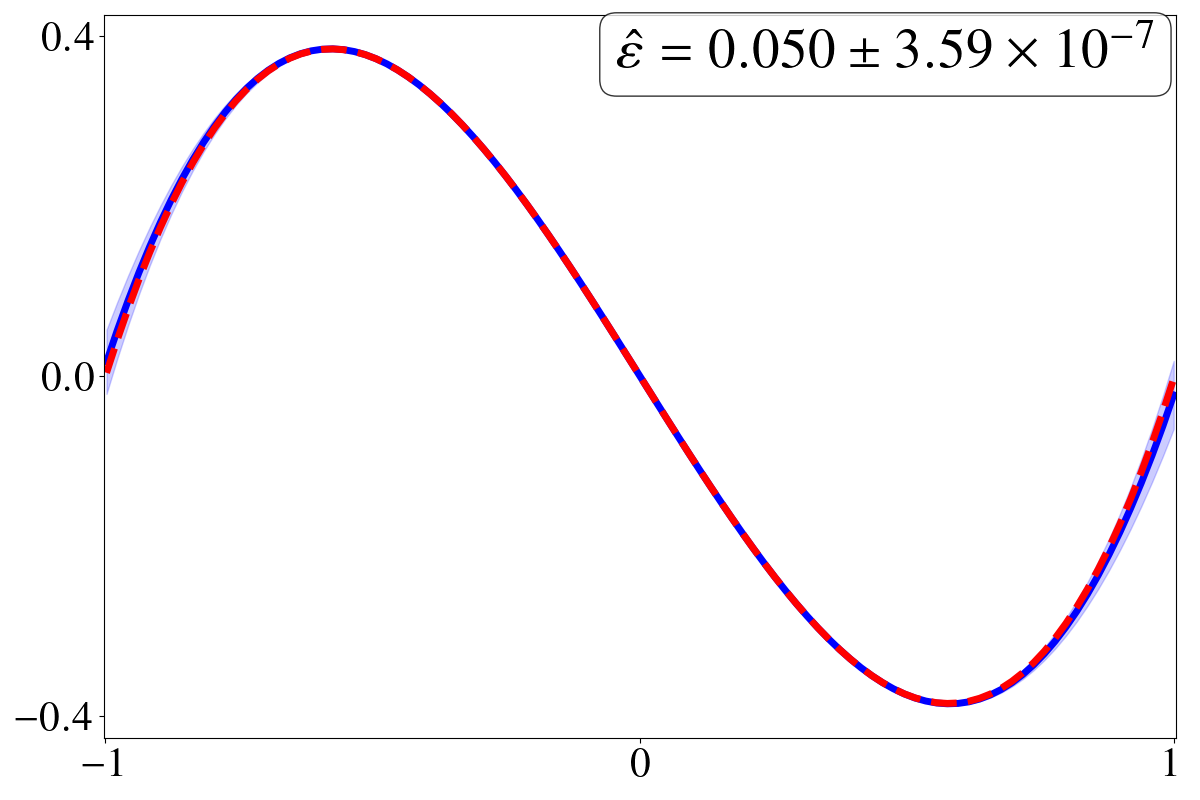}
    \end{subfigure}
    \begin{subfigure}[b]{0.49\textwidth}
        \includegraphics[width=\textwidth]{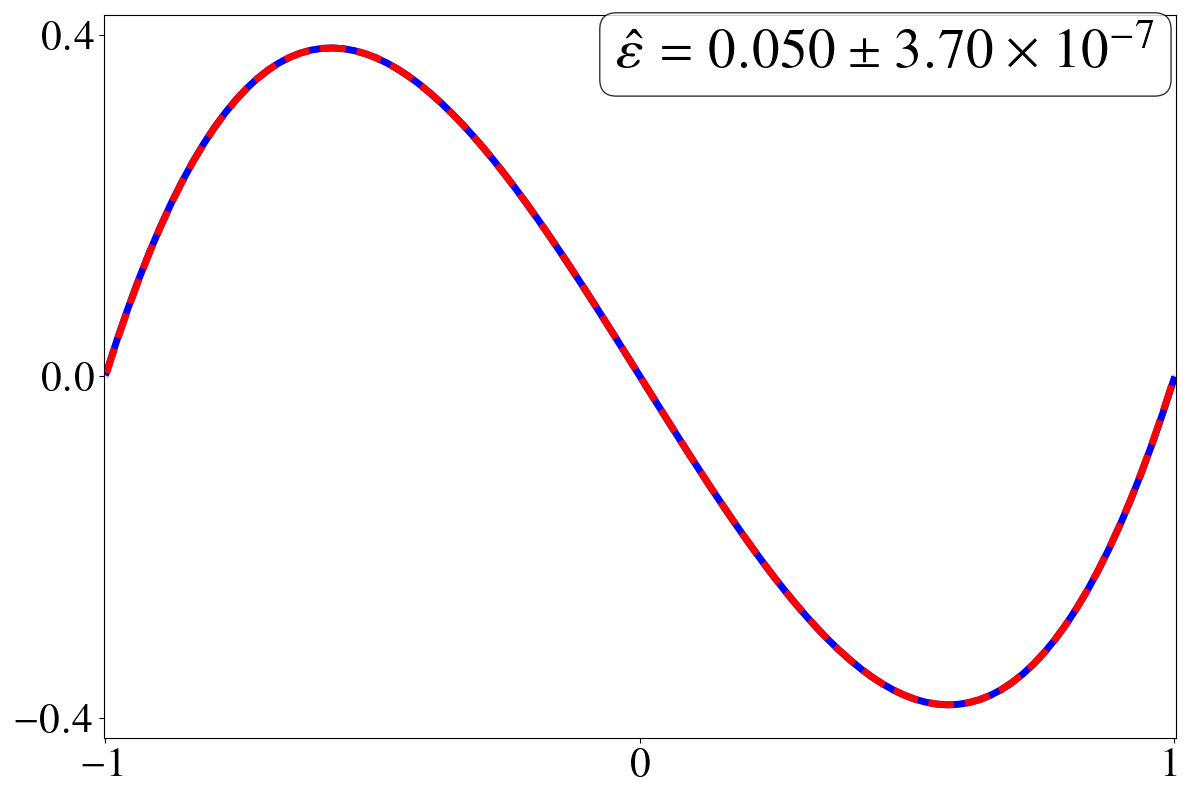}
    \end{subfigure} \\
    \begin{subfigure}[b]{0.49\textwidth}
        \includegraphics[width=\textwidth]{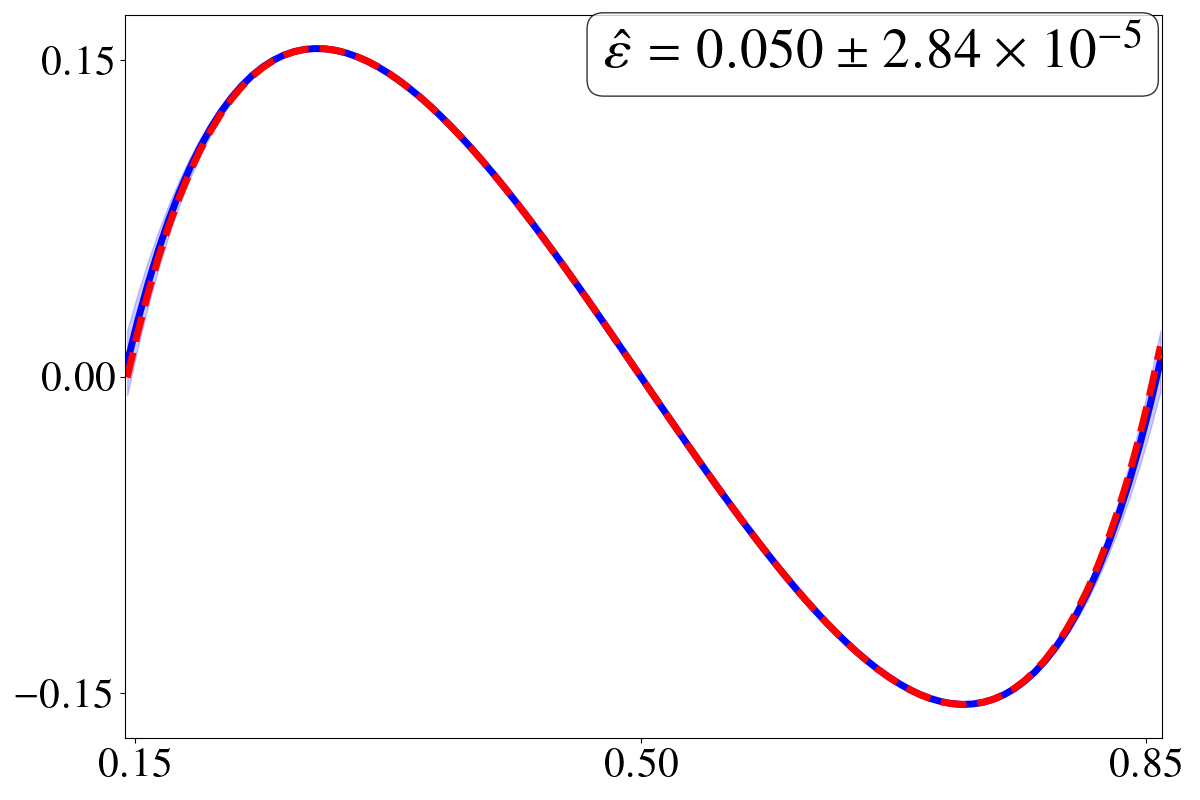}
    \end{subfigure}
    \begin{subfigure}[b]{0.49\textwidth}
        \includegraphics[width=\textwidth]{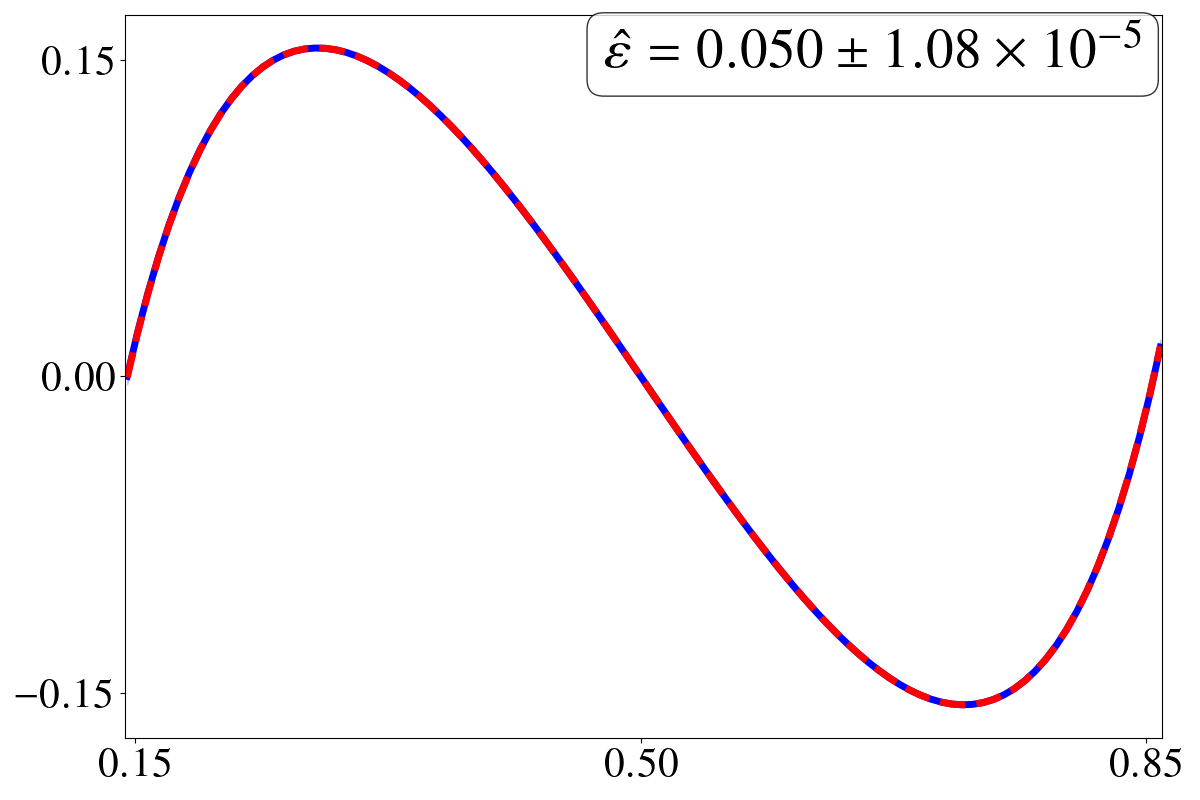}
    \end{subfigure}
    \caption{{\it Bulk chemical potential predictions from $100$ seeds on Cahn-Hilliard data for $N = 1,2$. The upper panel corresponds to the polynomial free energy~$(\ref{eqhf1})$ and the lower panel to the logarithmic free energy~$(\ref{eqhf2})$.}}
    \label{fig: noiseless results}
\end{figure}

\begin{figure}[tb]
    \centering
    \begin{subfigure}[b]{0.49\textwidth}
        \includegraphics[width=\textwidth]{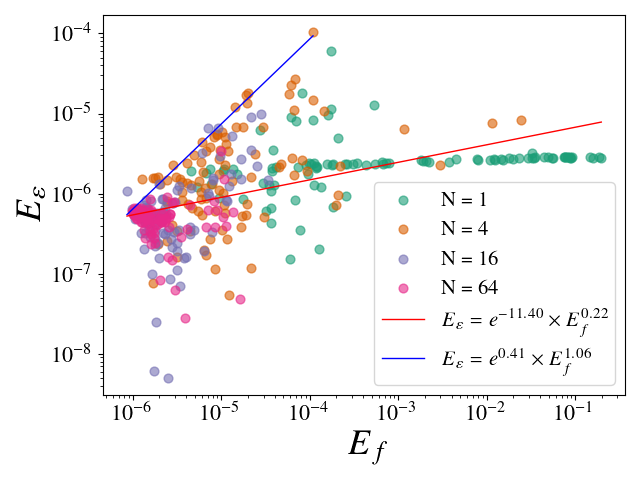}
    \end{subfigure}
    \begin{subfigure}[b]{0.49\textwidth}
        \includegraphics[width=\textwidth]{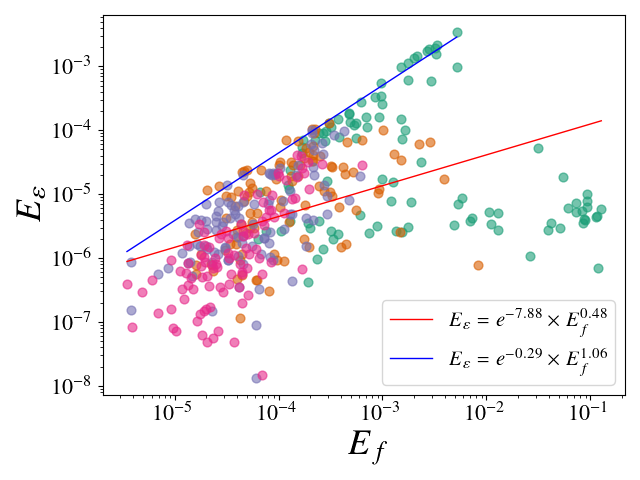}
    \end{subfigure}
    \caption{{\it Scatter graph showing the errors $E_f$ and $E_\varepsilon$ from $100$ seeds for various $N$. The left panel corresponds to the polynomial free energy~$(\ref{eqhf1})$ and the right panel to the logarithmix free energy~$(\ref{eqhf2})$. The red line is the best fit solution of all points, while the blue line represents the best fit solution of the $40$ points with the largest $E_\varepsilon / E_f$.}}
    \label{fig: noiseless E_f vs E_eps}
\end{figure}

\begin{figure}[tb]
    \centering
    \begin{subfigure}[b]{0.49\textwidth}
        \includegraphics[width=\textwidth]{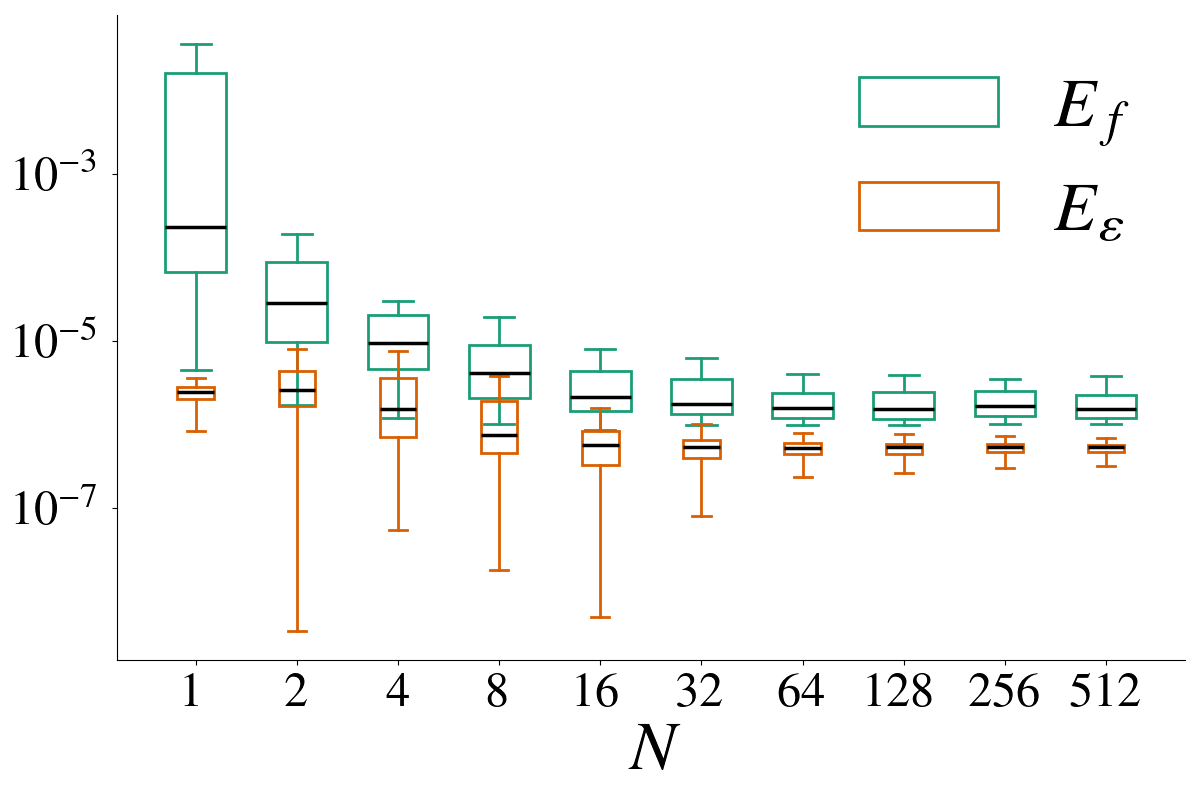}
    \end{subfigure}
    \begin{subfigure}[b]{0.49\textwidth}
        \includegraphics[width=\textwidth]{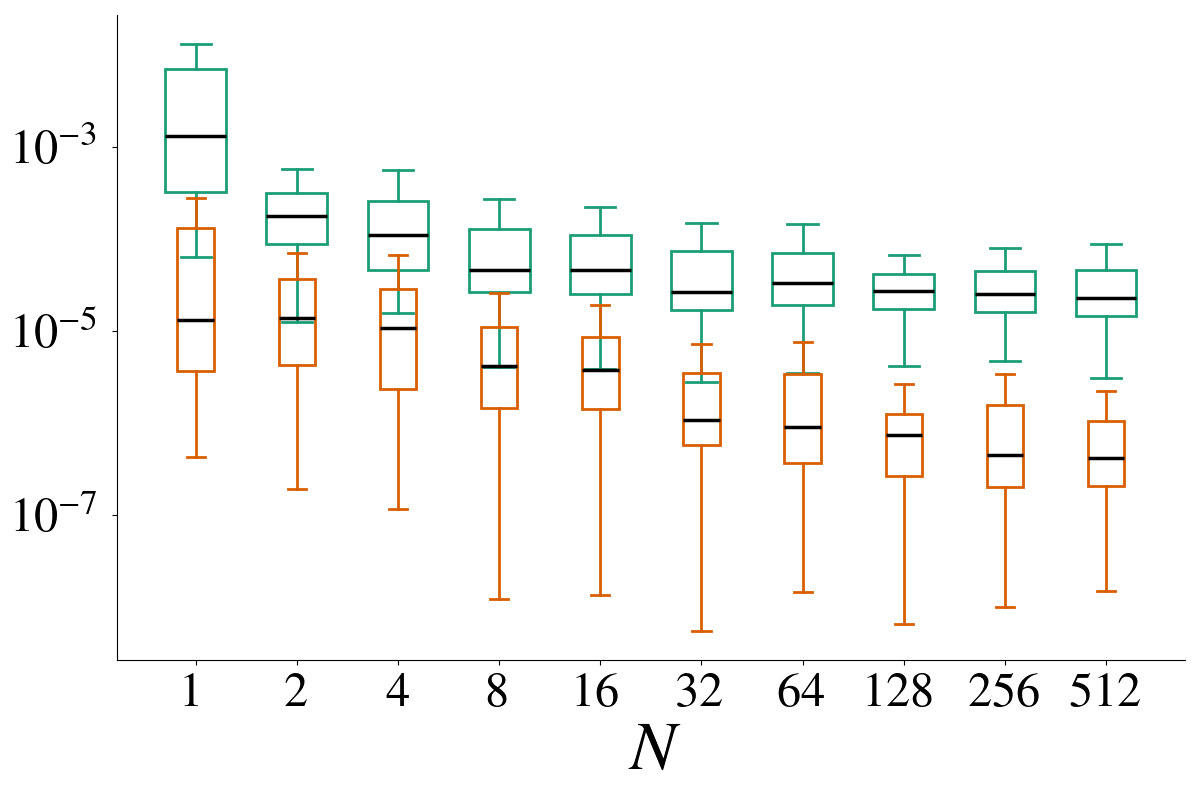}
    \end{subfigure}
    \caption{{\it Bar charts showing the errors $E_f$ and $E_\varepsilon$ from $100$ seeds as we increase $N$. The left panel corresponds to the polynomial free energy~$(\ref{eqhf1})$ and the right panel to the logarithmic free energy~$(\ref{eqhf2})$.}}
    \label{fig: noiseless E_f & E_eps vs N}
\end{figure}

Figure \ref{fig: noiseless E_f vs E_eps} quantifies the reconstruction accuracy across the 100 independent training runs by showing the joint distribution of the relative $L^2$ error in the bulk chemical potential, $E_f$, and the relative error in the interfacial thickness parameter, $E_\varepsilon$, for varying numbers of snapshot pairs $N$. Across both models, the method performs reliably even in low-data regimes. For $N=1$, a substantial proportion of runs already achieve small errors in both $E_f$ and $E_\varepsilon$, confirming that the extended SPINN framework can recover accurate constitutive information from minimal temporal input. As $N$ increases, the distribution of errors contracts markedly and shifts towards the lower-left corner of the plots, demonstrating systematic improvement in both the functional and parametric reconstruction. The reduction in spread with increasing $N$ indicates improved robustness with respect to the data points chosen, initialisation of the bulk free energy network and general optimisation variability up to $N=64$. In particular, the variance across seeds decreased significantly for larger $N$, while the mean error continues to decline. The behaviour is consistent across both free energy models.

Figure \ref{fig: noiseless E_f & E_eps vs N} initially shows that reconstruction accuracy improves substantially and variance decreases as more data points are added. However, increasing the number of data points beyond $N=64$ does not yield significant improvements in reconstruction quality. This suggests a limit on the number of additional data points required to achieve our best possible results. Overall, these results demonstrate that the proposed method achieves accurate and stable recovery of both the bulk chemical potential and interfacial thickness parameter on noiseless data, with performance improving predictably as additional snapshot information is provided.

\subsection{Application on noisy data}
\label{subsec: noise robustness}

\begin{figure}[tb]
    \centering

    \begin{subfigure}{0.32\textwidth}
        \centering
        \includegraphics[width=\linewidth]{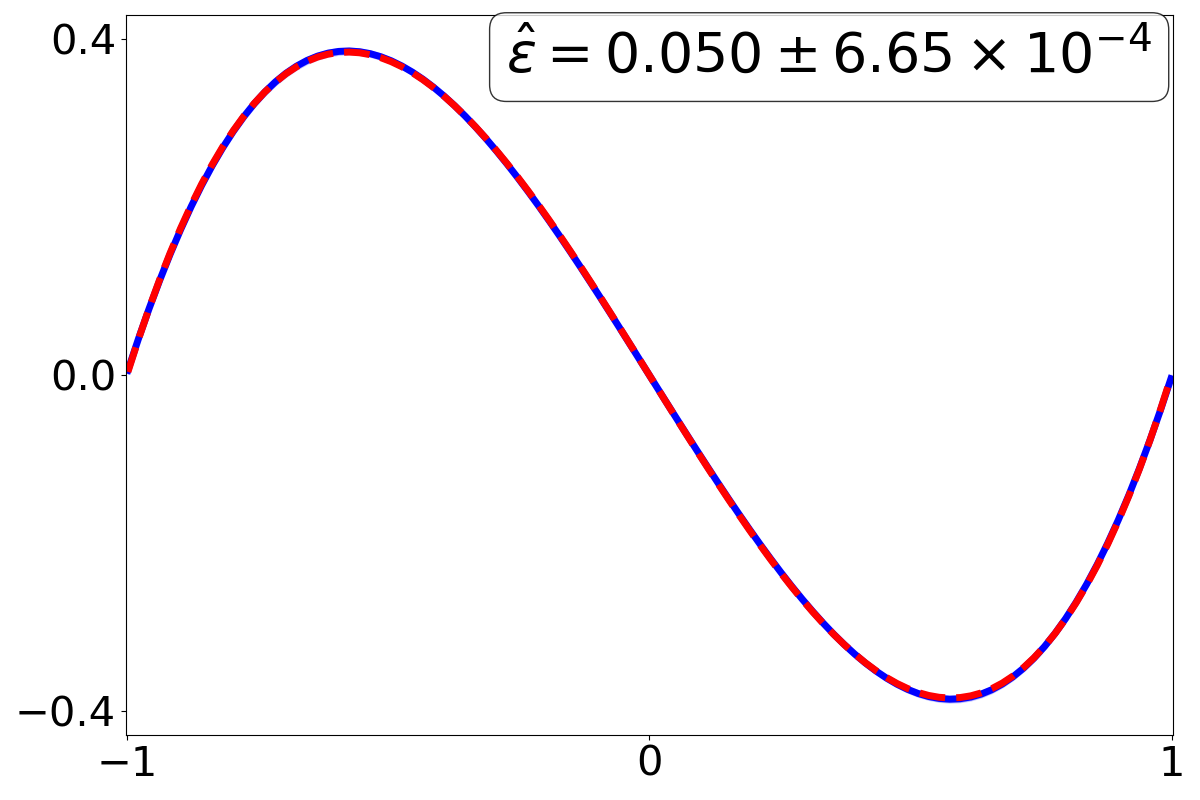}
    \end{subfigure}
    \hfill
    \begin{subfigure}{0.32\textwidth}
        \centering
        \includegraphics[width=\linewidth]{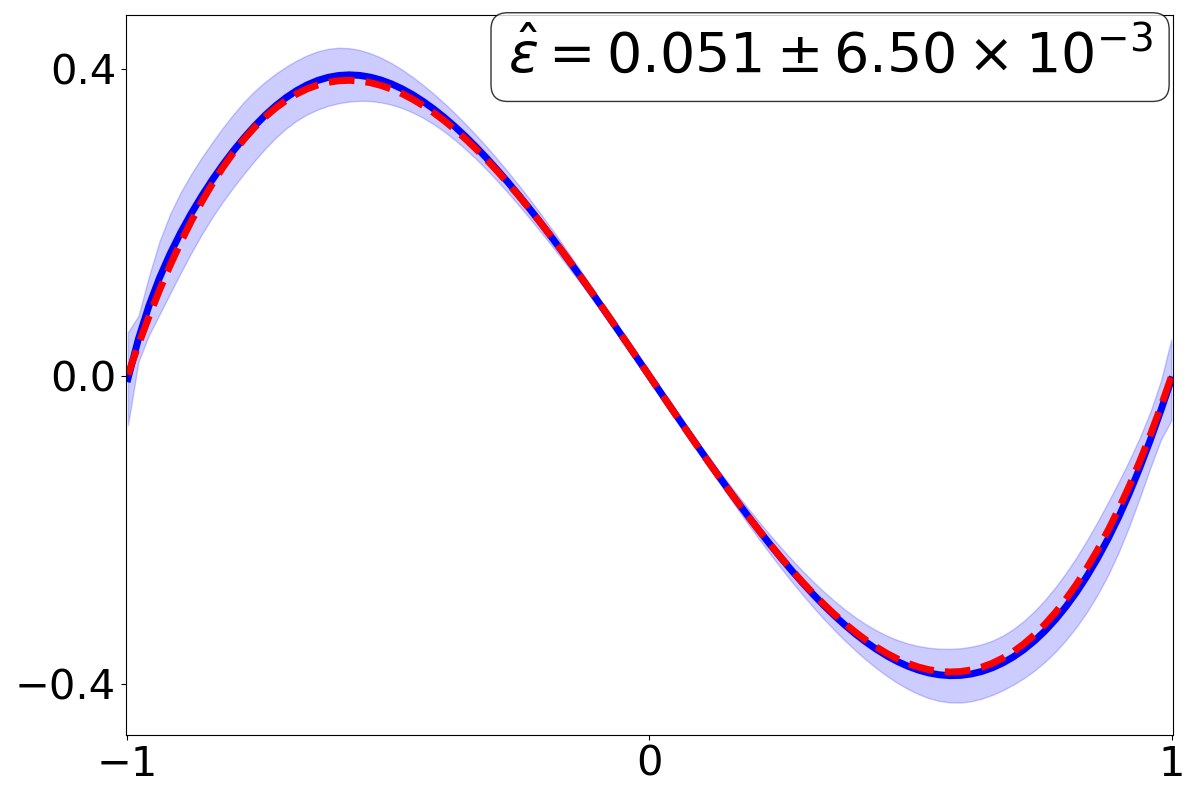}
    \end{subfigure}
    \hfill
    \begin{subfigure}{0.32\textwidth}
        \centering
        \includegraphics[width=\linewidth]{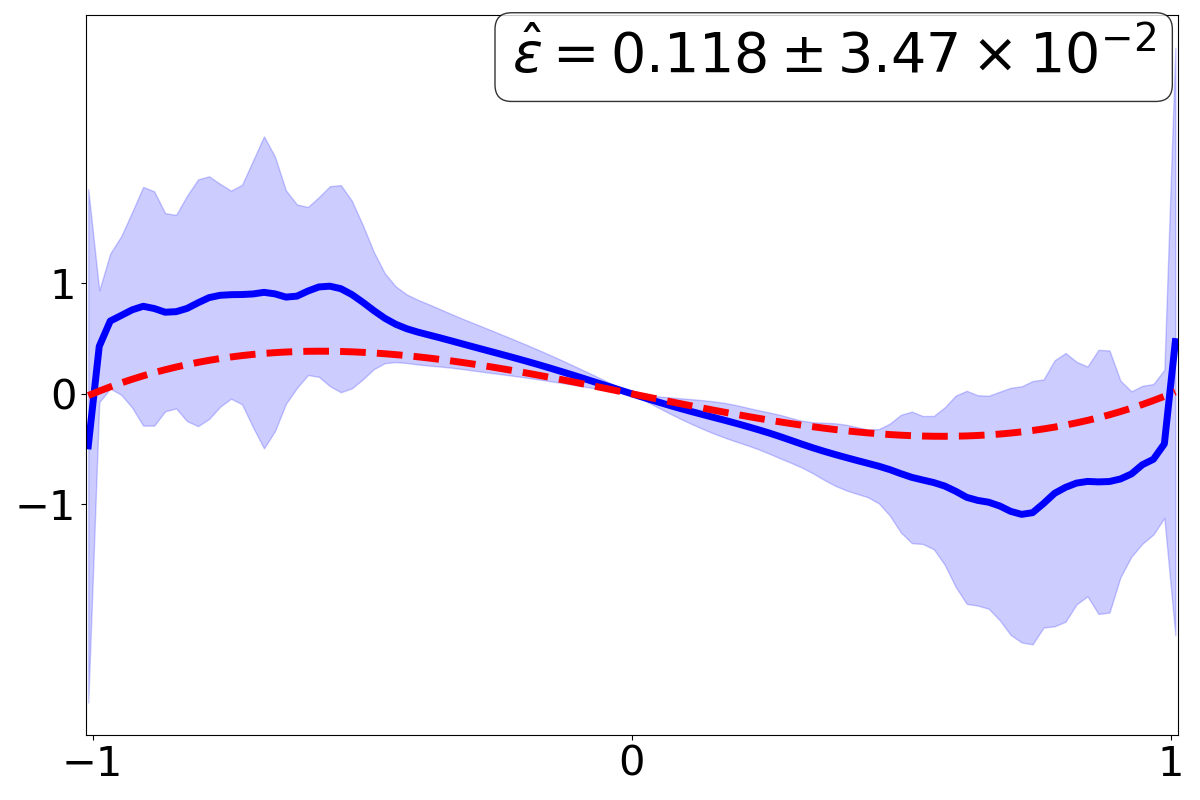}
    \end{subfigure}
    \\
        \begin{subfigure}{0.32\textwidth}
        \centering
        \includegraphics[width=\linewidth]{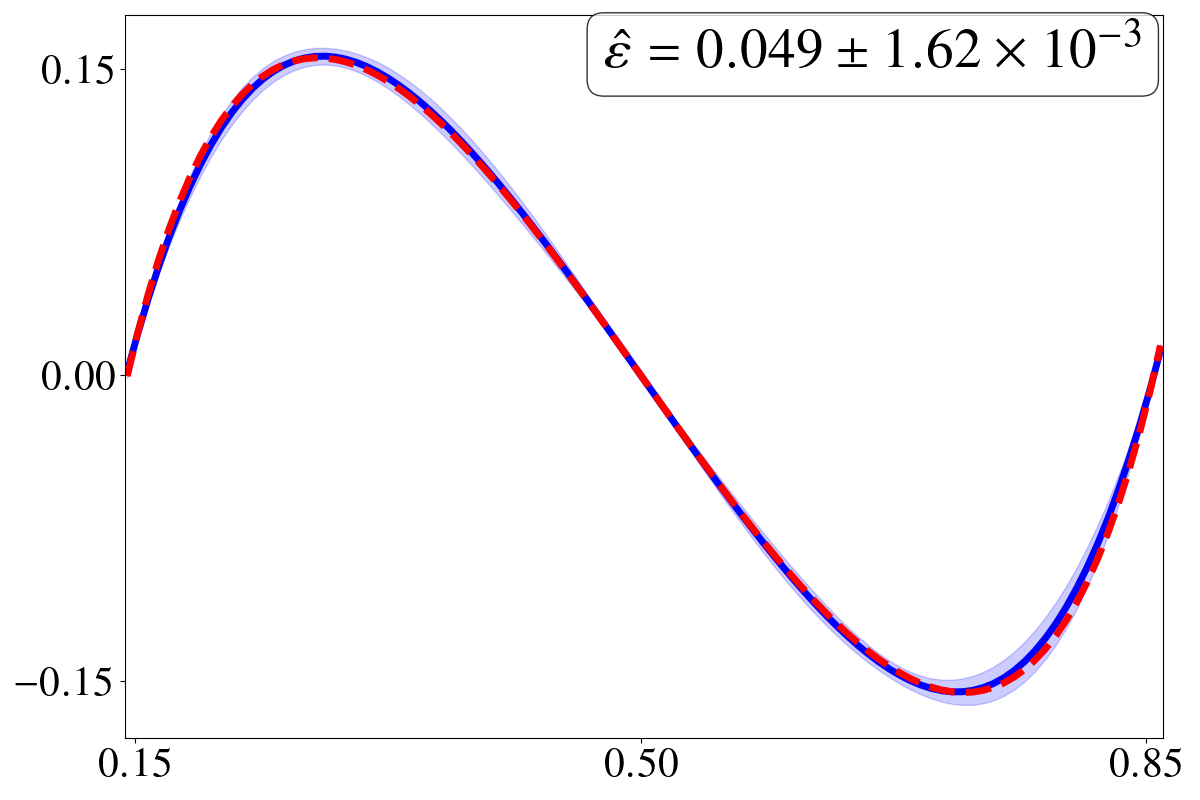}
        \caption{$\delta_L = 10^{-4}, \quad C_0 = 10^{5}$}
    \end{subfigure}
    \hfill
    \begin{subfigure}{0.32\textwidth}
        \centering
        \includegraphics[width=\linewidth]{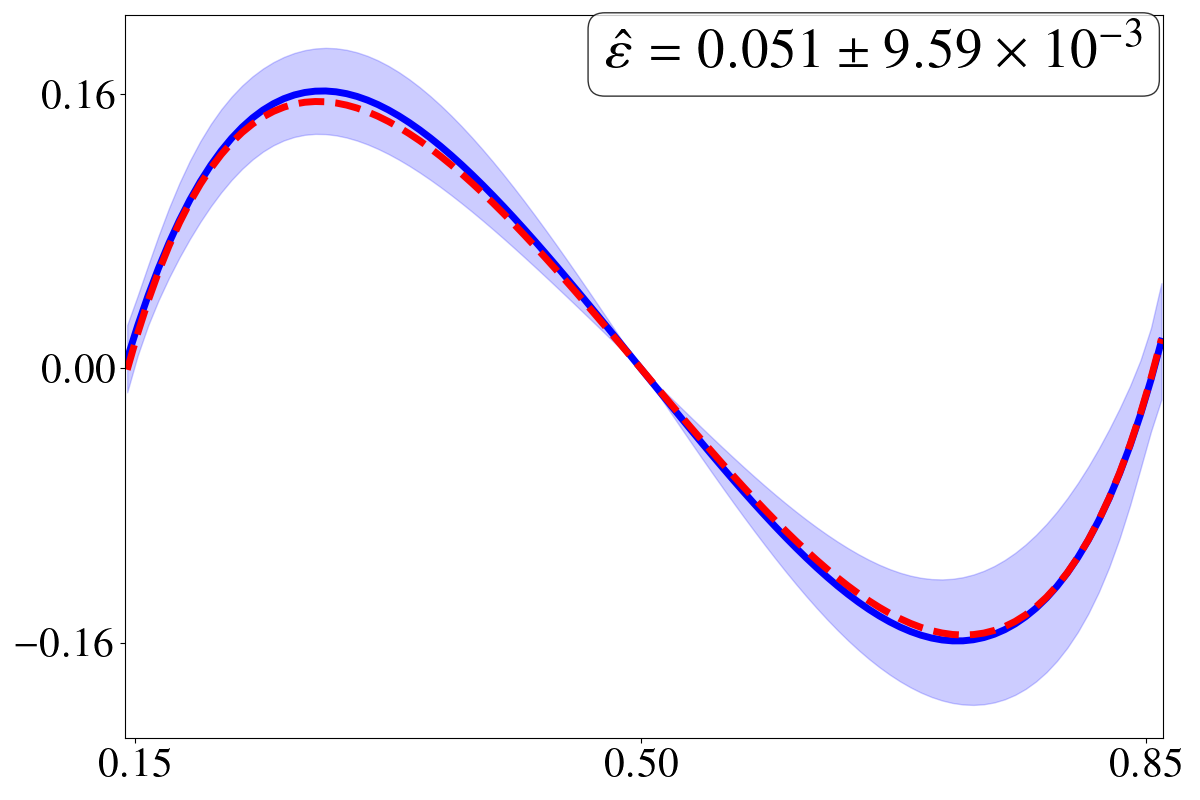}
        \caption{$\delta_L = 10^{-3}, \quad C_0 = 10^{3}$}
    \end{subfigure}
    \hfill
    \begin{subfigure}{0.32\textwidth}
        \centering
        \includegraphics[width=\linewidth]{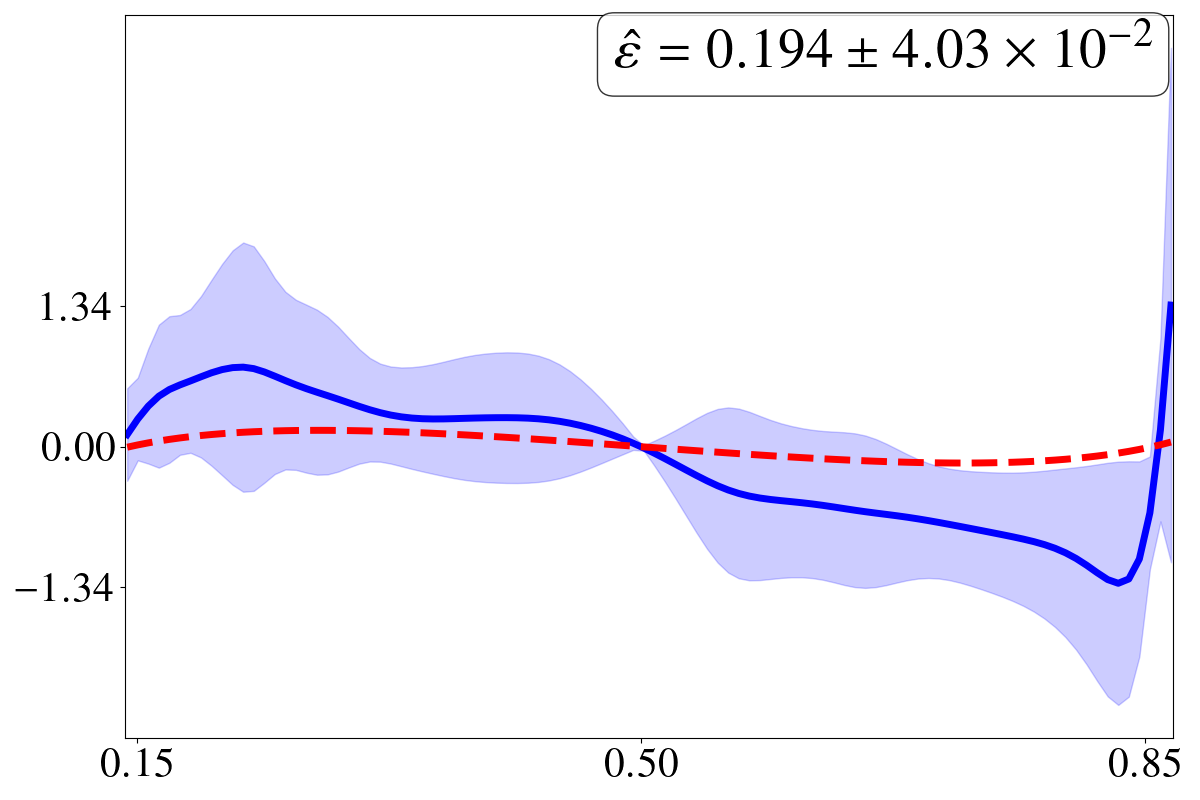}
        \caption{$\delta_L = 10^{-2}, \quad C_0 = 10^{1}$}
    \end{subfigure}

\caption{{\it Bulk chemical potential predictions from $100$ training runs on Cahn-Hilliard data with various noise levels and $N = 16$. The first row corresponds to the polynomial bulk chemical potential~$(\ref{eqhf1})$, and the second row to the logarithmic bulk chemical potential~$(\ref{eqhf2})$.}}
\label{fig: f robustness}
\end{figure}

To assess robustness, we next trained the model on noisy data. We introduce multiplicative noise to the data pre-training, where noise for each spatio-temporal data point is sampled from the uniform distribution $[-\delta_L, \delta_L]$. Note that, in contrast to the noiseless experiments, the optimisation strategy was modified for the noisy data regime. When observational noise is present, aggressively scaling the loss can amplify fluctuations in the residual and encourage overfitting to noise rather than recovery of the underlying constitutive structure. For this reason, we employed smaller loss multipliers specified, $C_0$, in Figure \ref{fig: f robustness}, thereby moderating gradient magnitudes and reducing sensitivity to stochastic perturbations in the data. Furthermore, we did not apply a subsequent L-BFGS refinement stage. While L-BFGS is highly effective for smooth, well-conditioned objectives, its quasi-Newton updates can become unstable or overly sensitive to noise, leading to overfitting and degraded results. Instead, we relied solely on Adam, which provides more robust stochastic updates and implicitly regularises the optimisation trajectory. These adjustments are made to ensure that the learned bulk chemical potential and interfacial thickness parameter reflect the dominant deterministic structure of the data rather than fitting to high-frequency noise components.

\begin{figure}[tb]
    \centering
    \begin{subfigure}[b]{0.49\textwidth}
        \includegraphics[width=\textwidth]{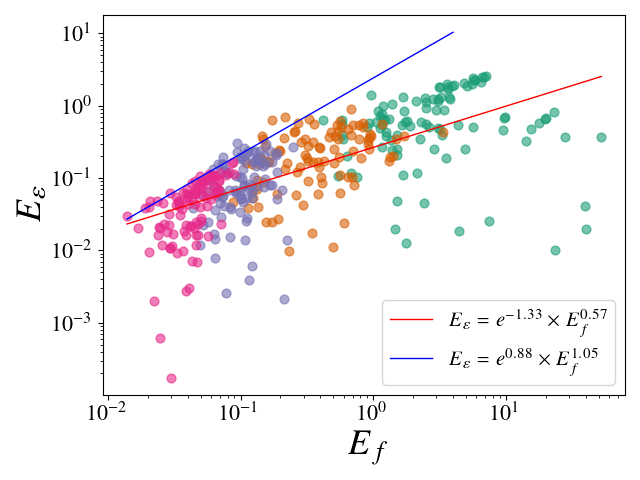}
    \end{subfigure}
    \begin{subfigure}[b]{0.49\textwidth}
        \includegraphics[width=\textwidth]{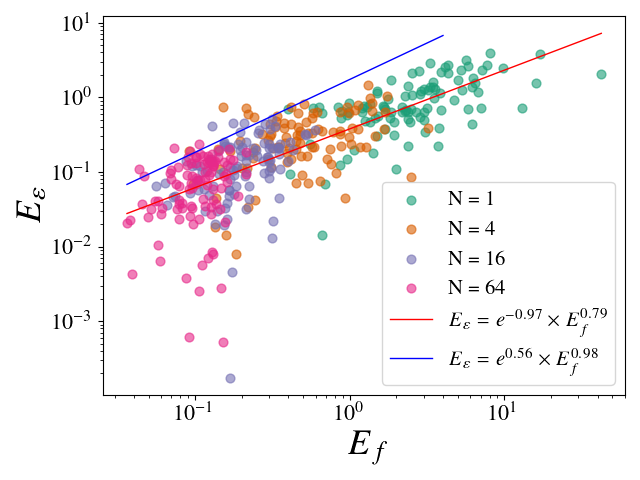}
    \end{subfigure}
    \caption{{\it Scatter graphs showing the errors $E_f$ and $E_\varepsilon$ from $100$ seeds for various $N$ from data with applied noise of amplitude $\delta_L=10^{-3}$ and $C_0 = 10^{3}$. The left graph shows results corresponding to the polynomial free energy~$(\ref{eqhf1})$, while the right graph shows results for the logarithmic free energy~$(\ref{eqhf2})$. The red line is the best fit solution of all points, while the blue line represents the best fit solution of the 40 points with the largest $E_\varepsilon / E_f$.}}
    \label{fig: noisy E_f vs E_eps}
\end{figure}

\begin{figure}[tb]
    \centering
    \begin{subfigure}[b]{0.49\textwidth}
        \includegraphics[width=\textwidth]{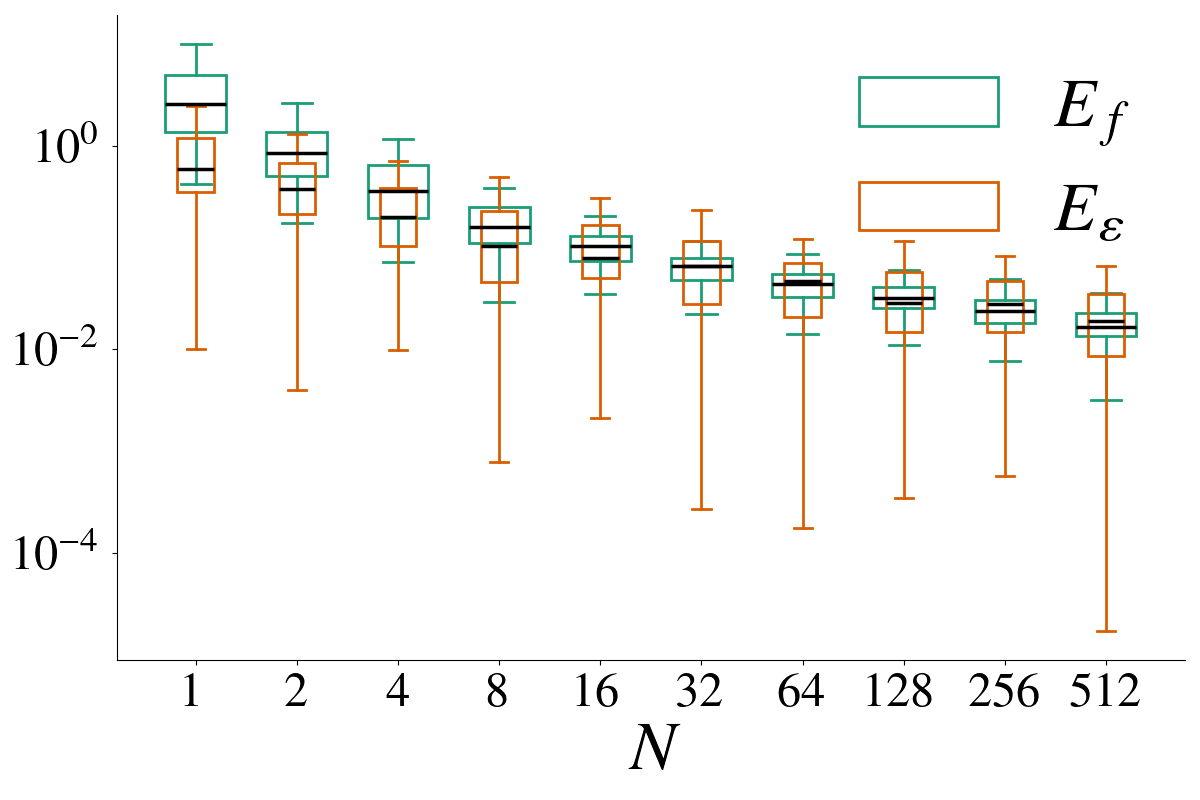}
    \end{subfigure}
    \begin{subfigure}[b]{0.49\textwidth}
        \includegraphics[width=\textwidth]{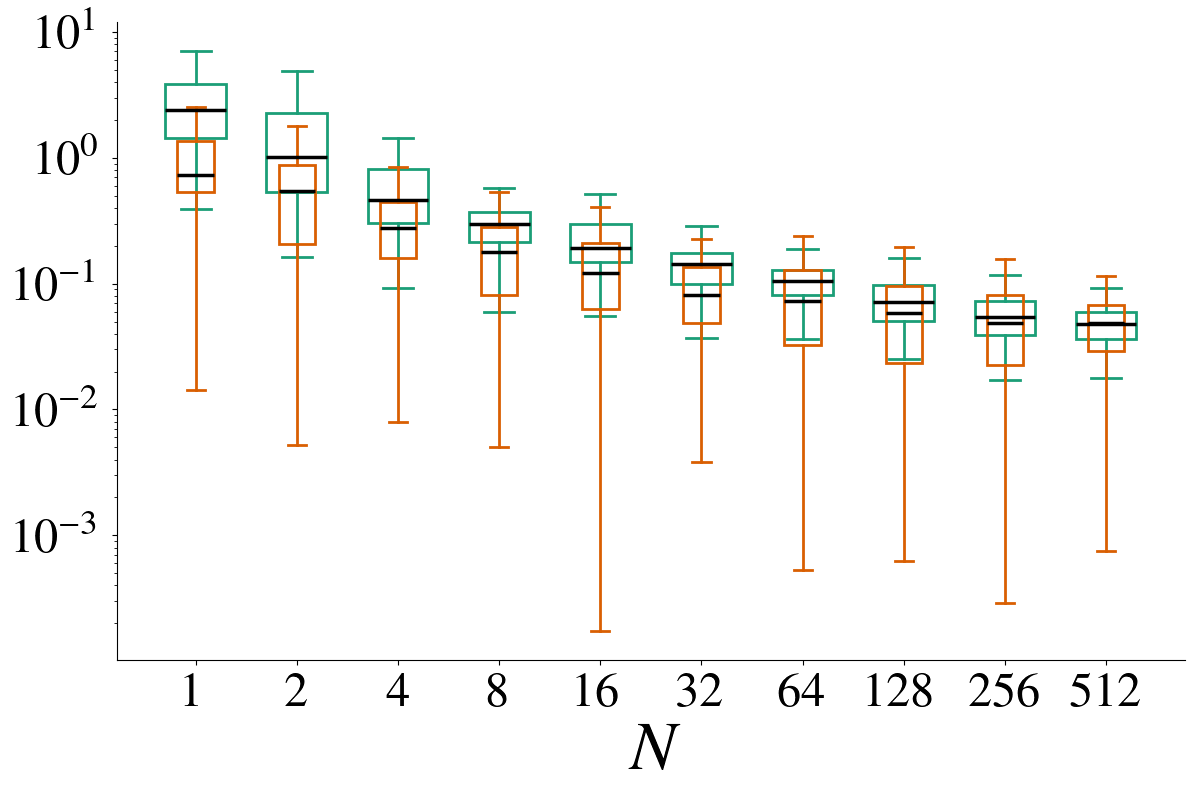}
    \end{subfigure}
    \caption{{\it Bar charts showing the errors $E_f$ and $E_\varepsilon$ from $100$ seeds as we increase $N$ trained on data with $\delta_L = 10^{-3}$. The left panel corresponds to the polynomial free energy~$(\ref{eqhf1})$ and the right panel to the logarithmic free energy~$(\ref{eqhf2})$.}}
    \label{fig: noisy E_f & E_eps vs N}
\end{figure}

Figure \ref{fig: noisy E_f vs E_eps} shows the joint distribution of relative errors $E_f$ and $E_\varepsilon$ across 100 independent training runs when a multiplicative noise amplitude of $\delta_L = 10^{-3}$ is added to the data. The loss multiplier is reduced to $C_0 = 10^{3}$, and optimisation is performed using Adam alone. As expected, the errors $E_f$ and $E_\varepsilon$ are significantly higher than in the noiseless case. On an absolute scale, the spread in errors across seeds is more pronounced. However, after adjusting for scale, we observe a similar spread to that in the noiseless case. For small $N$, the errors in both bulk chemical potential and the interfacial thickness parameter exhibit poor reconstruction (greater than $\mathcal{O}(1)$ errors for $N=1$) and a substantial variability. This shows that, unlike the noiseless case, limited noisy data impairs the method's ability to accurately and consistently reconstruct the free energy function and the interfacial thickness parameter.

Nevertheless, the method remains stable: even in the presence of noise, a clear reduction in both $E_f$ and $E_\varepsilon$ is observed as $N$ increases. The error clouds systematically shift to the lower-left region of the plots and become more concentrated for larger $N$, demonstrating that additional snapshot pairs improve both accuracy and robustness. This is consistent with what we would expect for noisy data, since each snapshot pair contains both the deterministic dynamical information (which we seek) and stochastic perturbations. With only a few snapshot pairs, the optimisation process may partially fit these perturbations, leading to variance across seeds. As $N$ increases, however, the influence of noise averages out across multiple temporal constraints, while the deterministic behaviour of the governing equations remains consistent. Consequently, the inverse problem becomes less sensitive to noise, and the learned quantities align more closely to their true value. In a sense, the additional snapshot pairs can be seen as an implicit regularisation mechanism, improving statistical stability without modifying the deterministic features. This behaviour is consistent between both the polynomial \eqref{eqhf1} and logarithmic \eqref{eqhf2} free energy models, although the latter again exhibits slightly larger variability. Figure \ref{fig: noisy E_f & E_eps vs N} shows how the trend of increasing $N$ continues to produce more accurate reconstructions. Unlike in the smooth data case, we continue to observe improved results for datasets with more than $N=64$ data points.

The blue line in Figures \ref{fig: noiseless E_f vs E_eps} and \ref{fig: noisy E_f vs E_eps} represent the linear fit to the $10\%$ of points with the largest $E_\varepsilon/E_f$, thereby approximating the upper envelope of the data. This reveals an effective limiting relation $E_\varepsilon \approx E_f + C$, suggesting that $E_\varepsilon$ sets a baseline error scale. We suggest that points lying close to this line represent cases where the framework is operating close to its best possible performance, whereas the points lying below this bound exhibit additional degradation in the solution for $E_f$. Equivalently, we can view the blue line as the worst possible performance of $E_\varepsilon$ for a given $E_f$.

Figure \ref{fig: f robustness} illustrates the effect of increasing $\delta_L$ on the reconstruction of the bulk chemical potential $\hat{f}$ and the associated interfacial thickness parameter $\hat{\varepsilon}$ with the snapshot pairs fixed at $N = 16$. For $\delta_L = 10^{-4}$, the ensemble means are nearly indistinguishable from the ground truth and variability across seeds is minimal. For $\delta_L = 10^{-3}$, the method continues to recover the correct functional structure. We observe an increased (but controlled) variance and a reasonably accurate estimate of the interfacial thickness parameter. The accurate mean and smooth curve suggest that the deviations are primarily stochastic rather than systematic at this noise level. For $N_L = 10^{-2}$, error increases with a visible bias in the learned function and thickness parameter.

\subsection{Architectural sensitivity}
\label{subsec: architectural sensitivity}

We evaluated how architectural choices influence the accuracy of the reconstructed bulk chemical potential and interfacial thickness parameter. Tables \ref{table: network architecture - poly} and \ref{table: network architecture epsilon - poly} report the final $E_f$ and $E_\varepsilon$, respectively, after Adam + L-BFGS, averaged over 100 seeds, for a range of activation functions and network sizes.

Across all experiments, the choice of activation function had a markedly greater impact on reconstruction accuracy than network depth or width. In particular, networks with ReLU activations consistently yielded larger reconstruction errors for both the bulk chemical potential and the interfacial thickness parameter, by up to four orders of magnitude relative to smooth activation functions. In contrast, the smooth activations sigmoid, tanh, SiLU, and GELU yielded comparable, consistently low errors across all tested architectures, whereas ELU performed slightly worse. Increasing the number of hidden layers or the number of neurons per layer had little consistent effect on the final accuracy of the construction (in some cases, the deeper network appeared to be a hindrance), suggesting that the inverse problem is not strongly capacity-limited. Instead, accurate reconstruction appears to rely primarily on the smoothness and differentiability properties of the activation function, rather than on increased network expressivity.

%\textcolor{red}{Table \ref{table: network architecture - noisey} tests the various combinations of network architectures and activation functions on noisy data. For these tests, a multiplicative noise of $N_L = 10^{-4}$ was applied to the data set before running the SPINNs algorithm. Some activation functions handle the additional noise better than others. Overall, on noiseless data, using a tanh activation function appears optimal. However, once we have noisy data, it is clear that the sigmoid and SiLU activation functions provide a superior reconstruction of the function $f(\phi)$. Table \ref{table: network architecture epsilon - poly} shows the final normalised $L^2$ error for the reconstructed $\varepsilon$. The error in the reconstructed $\varepsilon$ is consistently an order of magnitude less than the error in the reconstructed $f(\phi)$.  Based on the performance in our noiseless and noisy tests, we recommend a network architecture with 2 hidden layers, each with 20 neurons and a SiLU activation function. Moving forward, we use this combination.}

\begin{table}[tb]
\centering
\begin{tabular}{|c||c|c|c|}
\hline
\diagbox[width=15em]{[$n_1$, $n_2$, $\dots$, $n_\ell$]}{$\omega$}
&
\parbox{18mm}{\centering sigmoid $\sigma$}
&
\parbox{15mm}{\centering tanh}
&
\parbox{13mm}{\centering ReLU}
\\
\hline
\hline
\rule{0pt}{4mm} [10, 10] & $3.71 \times10^{-6}$ & $3.71 \times10^{-6}$ & $5.13 \times10^{-2}$ \\
\hline
\rule{0pt}{4mm} [20, 20] & $5.38 \times10^{-6}$ & $4.89 \times10^{-6}$ & $1.78 \times10^{-2}$ \\
\hline
\rule{0pt}{4mm} [32, 32] & $4.72 \times10^{-6}$ & $4.05 \times10^{-6}$ & $1.33 \times10^{-2}$ \\
\hline
\rule{0pt}{4mm} [32, 32, 32, 32] & $3.44 \times10^{-6}$ & $7.97 \times10^{-6}$ & $1.73 \times10^{-2}$ \\
\hline
\rule{0pt}{4mm} [64, 64, 64, 64] & $6.06 \times10^{-6}$ & $1.40 \times 10^{-5}$ & $1.17 \times10^{-2}$ \\
\hline
\hline

\diagbox[width=15em]{[$n_1$, $n_2$, $\dots$, $n_\ell$]}{$\omega$}
&
\parbox{18mm}{\centering SiLU}
&
\parbox{15mm}{\centering ELU}
&
\parbox{13mm}{\centering GELU}
\\
\hline
\hline
\rule{0pt}{4mm} [10, 10] & $4.11 \times10^{-6}$ & $2.70 \times10^{-4}$ & $4.92 \times10^{-6}$ \\
\hline
\rule{0pt}{4mm} [20, 20] & $4.63 \times10^{-6}$ & $2.21 \times10^{-4}$ & $ 3.82 \times10^{-6}$ \\
\hline
\rule{0pt}{4mm} [32, 32] & $5.55 \times10^{-6}$ & $2.02 \times10^{-4}$ & $3.79 \times10^{-6}$ \\
\hline
\rule{0pt}{4mm} [32, 32, 32, 32] & $3.44 \times10^{-6}$ & $2.03 \times10^{-4}$ & $ 2.12 \times10^{-5}$ \\
\hline
\rule{0pt}{4mm} [64, 64, 64, 64] & $3.71 \times10^{-6}$ & $1.71 \times10^{-4}$ & $1.07 \times10^{-5}$ \\
\hline
\end{tabular}
\caption{{\it $E_f$ on the polynomial free energy model~$(\ref{eqhf1})$ for different network sizes and activation functions over 100 initial seeds with $N = 16$.}}
\label{table: network architecture - poly}
\end{table}

\begin{table}[tb]
\centering
\begin{tabular}{|c||c|c|c|}
\hline
\diagbox[width=15em]{[$n_1$, $n_2$, $\dots$, $n_\ell$]}{$\omega$}
&
\parbox{18mm}{\centering sigmoid $\sigma$}
&
\parbox{15mm}{\centering tanh}
&
\parbox{13mm}{\centering ReLU}
\\
\hline
\hline
\rule{0pt}{4mm} [10, 10] & $8.01 \times10^{-7}$ & $8.51 \times10^{-7}$ & $6.75 \times10^{-3}$ \\
\hline
\rule{0pt}{4mm} [20, 20] & $1.13 \times10^{-6}$ & $8.13 \times10^{-7}$ & $1.92 \times10^{-3}$ \\
\hline
\rule{0pt}{4mm} [32, 32] & $9.09 \times10^{-7}$ & $8.35 \times10^{-7}$ & $7.88 \times10^{-4}$ \\
\hline
\rule{0pt}{4mm} [32, 32, 32, 32] & $8.89 \times10^{-7}$ & $6.40 \times10^{-7}$ & $1.30 \times10^{-3}$ \\
\hline
\rule{0pt}{4mm} [64, 64, 64, 64] & $8.39 \times10^{-7}$ & $6.70 \times 10^{-7}$ & $7.79 \times10^{-4}$ \\
\hline
\hline

\diagbox[width=15em]{[$n_1$, $n_2$, $\dots$, $n_\ell$]}{$\omega$}
&
\parbox{18mm}{\centering SiLU}
&
\parbox{15mm}{\centering ELU}
&
\parbox{13mm}{\centering GELU}
\\
\hline
\hline
\rule{0pt}{4mm} [10, 10] & $8.73 \times10^{-7}$ & $8.26 \times10^{-6}$ & $1.31 \times10^{-6}$ \\
\hline
\rule{0pt}{4mm} [20, 20] & $1.03 \times10^{-6}$ & $6.84 \times10^{-6}$ & $ 1.02 \times10^{-6}$ \\
\hline
\rule{0pt}{4mm} [32, 32] & $1.44 \times10^{-6}$ & $5.31 \times10^{-6}$ & $9.25 \times10^{-6}$ \\
\hline
\rule{0pt}{4mm} [32, 32, 32, 32] & $6.68 \times10^{-7}$ & $4.55 \times10^{-6}$ & $7.17 \times10^{-7}$ \\
\hline
\rule{0pt}{4mm} [64, 64, 64, 64] & $6.69 \times10^{-7}$ & $4.29 \times10^{-6}$ & $7.58 \times10^{-7}$ \\
\hline
\end{tabular}
\caption{{\it $E_\varepsilon$ on the polynomial free energy model~$(\ref{eqhf1})$ for different network sizes and activation functions over $100$ initial seeds with $N = 16$.}}
\label{table: network architecture epsilon - poly}
\end{table}

\subsection{Learning rate sensitivity}
\label{subsec: learning rate sensitivity}

\begin{figure}[tb]
\centering
\includegraphics[width=\linewidth]{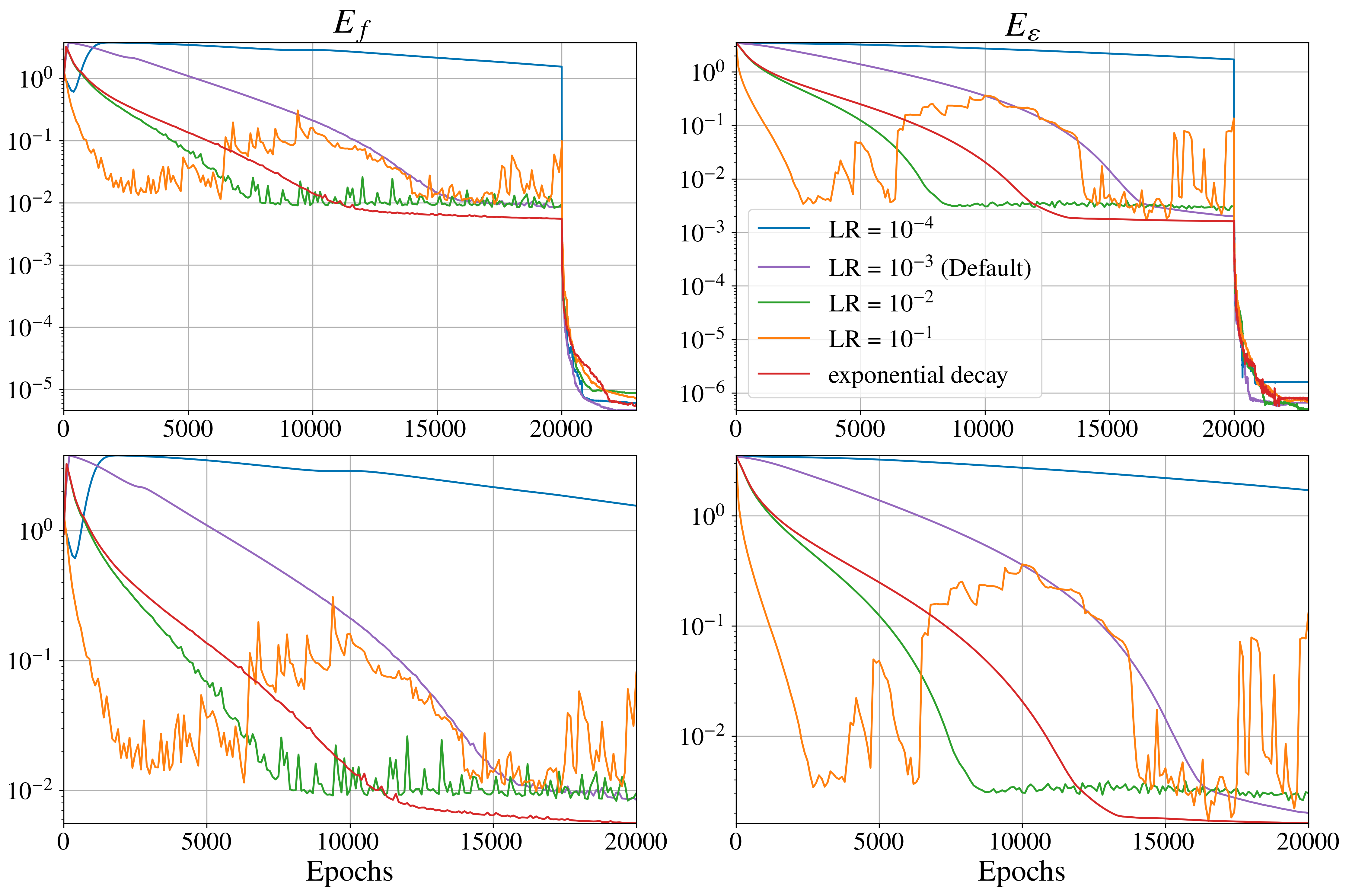}
\caption{{\it Graphs showing the mean trajectory of $E_f$ and $E_\varepsilon$ of $10$ seeds throughout the training process for several learning rates, as well as an exponentially decaying learning rate for $N = 16$. The top row shows the trajectory for Adam and L-BFGS training, while the bottom row shows only Adam training.}}
\label{fig: learning curves}
\end{figure}

Figure \ref{fig: learning curves} shows the mean $E_f$ and $E_\varepsilon$ throughout the network training process for several fixed learning rates along with an exponentially decaying learning rate. The method is somewhat sensitive to the choice of learning rate in Adam optimisation. Firstly, we observe that using a smaller learning rate yields smooth decay of the errors $E_f$ and $E_\varepsilon$, suggesting that these learning rates are well-conditioned in the loss landscape and that the optimisation proceeds through a stable basin rather than oscillatory or chaotic regimes. Conversely, larger learning rates lead to visibly unstable training dynamics and oscillatory curves. Choosing a learning rate that is too small ($10^{-4}$ in the graphs) leads to slow learning and a poor reconstruction after $20,000$ epochs, while choosing a learning rate that is too large ($10^{-1}$ in the graphs) results in Adam failing to converge, suggesting that it overshoots the actual minima. Comparing the learning rates that converge correctly in the Adam stage, we observe that although a learning rate of $10^{-2}$ converges faster than the default, it produces visible oscillations. 

To exploit the fast convergence achieved with a learning rate of $10^{-2}$ while mitigating instability in the learning curve, we tested an exponential learning rate initialised at $10^{-2}$ and decaying to $10^{-4}$. This resulted in a quicker convergence and lower errors at the end of Adam training. It is noted that once L-BFGS has been applied to noiseless data, it typically converges even with non-converging Adam learning rates. However, since we cannot rely on L-BFGS for noisy data, it was important to assess how the learning rates performed with Adam alone.

\section{Discussion}
\label{sec: conclusions}

In this work, we have developed the ESPINN framework for the inverse identification of constitutive structure in phase-field models. Building on the SPINN methodology of Zhao~\cite{zhao2021discoveringphasefieldmodels}, the proposed extension enables the simultaneous recovery of the bulk chemical potential and unknown gradient coefficients, including the interfacial thickness parameter $\varepsilon$, from dynamically evolving snapshot data. In the context of phase separation, these quantities determine the free-energy landscape and interfacial structure that govern the pattern formation and coarsening dynamics.

Our numerical experiments on the Cahn-Hilliard model~(\ref{cahnhilliardeq}) demonstrate that the ESPINN framework accurately reconstructs both the functional form of the bulk chemical potential and the associated interfacial thickness parameter in the noiseless regime. Even a single snapshot pair contains sufficient dynamical information to recover the free-energy landscape with additional temporal data, thereby systematically improving statistical stability and reducing variance across random seed initialisations. This highlights the method's ability to extract the information encoded in transient phase dynamics. In the presence of noise, reconstruction accuracy degrades, but in a controlled manner, and the method remains robust over a broad range of noise amplitudes. Increasing the number of snapshot pairs significantly mitigates the impact of noise, effectively averaging out the stochastic perturbations while preserving the deterministic structure~\cite{Erban:2020:SMR}. This indicates that the transient dynamical data provide a reliable basis for learning the constitutive information, even when observations are imperfect.

PINNs have traditionally been used not only for inverse problems, where unknown parameters in PDEs are inferred from time-series data~\cite{Erban:2026:NNL,zhao2021discoveringphasefieldmodels}, but also for solving PDEs with known governing equations. In this setting, an alternative approach for the Allen–Cahn and Cahn–Hilliard equations is based on using physics-informed neural operators (PINOs), which are capable of learning solution operators across a range of PDE settings~\cite{Gangmei:2025:LCA}. Although such methods typically require more training data, they could, in principle, also be adapted to inverse problems when trained on data from related trajectories.

In the presented numerical tests, attention was restricted to the one-dimensional Cahn-Hilliard model with constant mobility to isolate the core mechanisms relevant to inverse identification and coarse-graining. % We provide the equivalent results for the Allen-Cahn model~(\ref{allencahn}) in the {\it Supplementary Material}.
9This setting enables detailed numerical diagnostics and clear comparisons across different modelling and learning approaches, while avoiding the additional complexities of higher-dimensional geometries. Extensions to higher spatial dimensions, more general mobility laws, and direct application to experimental data are beyond the scope of the present work and are left for future investigation.

The ESPINN framework, therefore, offers a practical and data-efficient approach to inferring free-energy structure and interfacial parameters in continuum phase-field models. While the current study is restricted to one-dimensional models with constant mobility, the methodology naturally extends to higher-dimensional settings and more complex free-energy functionals. Such extensions are particularly relevant for coarse-grained models of biomolecular condensates, polymer blends and alloy systems, where accurate identification of the free-energy landscapes is central to predictive modelling.

\vskip 2mm

\noindent
{\bf Acknowledgements.} We thank Georg Meyerhofer, Giulia Celora and Ruth Baker for useful discussions.

\vskip 2mm

\noindent
{\bf Data Availability.}  In compliance with EPSRC’s open access initiative, the data in this paper is available
from: https://doi.org/10.5281/zenodo.20797058

\vskip 2mm

\noindent
{\bf Ethics.} This study did not involve human participants, animals, or sensitive data requiring
ethical approval or consent to participate.

\vskip 2mm

\noindent
{\bf Conflicts of interest.} The authors declare no conflicts of interest.

\bibliographystyle{siamplain}
\bibliography{references}
\end{document}